\documentclass[fleqn,usenatbib,useAMS]{mnras}
\usepackage{graphicx}
\usepackage{amsmath}
\usepackage{CJKutf8}
\usepackage{newtxtext,newtxmath}
\usepackage[T1]{fontenc}
\usepackage{xcolor}
\usepackage{xspace}
\usepackage{xurl}
\usepackage{array}
\usepackage{orcidlink}

\newcommand{\kms}{\mbox{km\,s$^{-1}$}}
\newcommand{\mjypbm}{\mbox{mJy\,beam$^{-1}$}}
\newcommand{\msol}{\mbox{$M_\odot$}}
\newcommand{\arcdeg}{\mbox{$^\circ$}}
\newcommand{\cc}{\mbox{cm$^{-3}$}}

\newcommand{\vlsr}{\mbox{$V_\textrm{LSR}$}}
\newcommand{\hii}{\mbox{H\textsc{ii}}}

\newcommand{\htcn}{\mbox{H$^{13}$CN}}
\newcommand{\htcop}{\mbox{H$^{13}$CO$^+$}}
\newcommand{\hcop}{\mbox{HCO$^+$}}
\newcommand{\hntc}{\mbox{HN$^{13}$C}}
\newcommand{\hcfn}{\mbox{HC$^{15}$N}}

\defcitealias{Longmore2026}{Paper~I}
\defcitealias{Ginsburg2026}{Paper~II}
\defcitealias{Walker2026}{Paper~III}
\defcitealias{Hsieh2026}{Paper~V}

\title[ACES INTERMEDIATE WIDTH SPECTRAL WINDOWS]{ALMA Central molecular zone Exploration Survey (ACES) IV:\\ Data of the two intermediate-width spectral windows}

\newcounter{affcounter}

\newcommand{\defaffiliationlabel}[1]{%
  \refstepcounter{affcounter}%
  \expandafter\xdef\csname #1\endcsname{\theaffcounter}%
}

\newcommand{\affref}[1]{$^{\csname #1\endcsname}$}

\newcommand{\affrefs}[1]{%
  $^{%
    \@for\@ref:=#1\do{%
      \@ref\@ifnextchar\@nil{}{,}%
    }%
  }$%
}

\newcommand{\affrefTwo}[2]{$^{\csname #1\endcsname,\csname #2\endcsname}$}
\newcommand{\affrefThree}[3]{$^{\csname #1\endcsname,\csname #2\endcsname,\csname #3\endcsname}$}
\newcommand{\affrefFour}[4]{$^{\csname #1\endcsname,\csname #2\endcsname,\csname #3\endcsname,\csname #4\endcsname}$}

\newcommand{\printaffiliation}[2]{%
  $^{\csname #1\endcsname}$#2\\%
}
\defaffiliationlabel{shao}
\defaffiliationlabel{naoc_key}
\defaffiliationlabel{ukarcnode}
\defaffiliationlabel{uflorida}
\defaffiliationlabel{eso}
\defaffiliationlabel{naoj}
\defaffiliationlabel{ice_csic}
\defaffiliationlabel{ieec}
\defaffiliationlabel{umd}
\defaffiliationlabel{mpe}
\defaffiliationlabel{kansas}
\defaffiliationlabel{ljmu}
\defaffiliationlabel{mpia}
\defaffiliationlabel{ari_heidelberg}
\defaffiliationlabel{COOL}
\defaffiliationlabel{eso_chile}
\defaffiliationlabel{jao}
\defaffiliationlabel{nanjing}
\defaffiliationlabel{nanjing_key}
\defaffiliationlabel{cfa}
\defaffiliationlabel{colorado}
\defaffiliationlabel{uconn}
\defaffiliationlabel{cab_csic}
\defaffiliationlabel{iaa_taipei}
\defaffiliationlabel{chalmers}
\defaffiliationlabel{clap}
\defaffiliationlabel{iop_epfl}
\defaffiliationlabel{oaq}
\defaffiliationlabel{inaf_arcetri}
\defaffiliationlabel{jbca}
\defaffiliationlabel{nrao}
\defaffiliationlabel{ita_heidelberg}
\defaffiliationlabel{anu}
\defaffiliationlabel{iaa_csic}
\defaffiliationlabel{ucn}
\defaffiliationlabel{cassaca}
\defaffiliationlabel{ias}
\defaffiliationlabel{ucl}
\defaffiliationlabel{izw_heidelberg}
\defaffiliationlabel{ipm}
\defaffiliationlabel{ulaserena}
\defaffiliationlabel{iff_csic}
\defaffiliationlabel{gbo}
\defaffiliationlabel{utokyo}
\defaffiliationlabel{umass}
\defaffiliationlabel{aberystwyth}
\defaffiliationlabel{kiaa_pku}
\defaffiliationlabel{pku_astro}
\defaffiliationlabel{ist}
\author[X.~Lu \& ACES Team]{Xing Lu,\affrefTwo{shao}{naoc_key}\orcidlink{0000-0003-2619-9305}
Daniel L.~Walker,\affref{ukarcnode}\thanks{E-mail: daniel.walker.astro@gmail.com}\orcidlink{0000-0001-7330-8856}
Adam Ginsburg,\affref{uflorida}\orcidlink{0000-0001-6431-9633}
Ashley~T.~Barnes,\affref{eso}\orcidlink{0000-0003-0410-4504}
Pei-Ying Hsieh,\affref{naoj}\orcidlink{0000-0001-9155-39}
\newauthor
\'Alvaro S\'anchez-Monge,\affrefTwo{ice_csic}{ieec}\orcidlink{0000-0002-3078-9482}
Savannah R. Gramze,\affref{uflorida}\orcidlink{0000-0002-1313-429X}
Nazar Budaiev,\affref{uflorida}\orcidlink{0000-0002-0533-8575}
Marc W.~Pound,\affref{umd}\orcidlink{0000-0002-7269-342X}
\newauthor
Jaime E. Pineda,\affref{mpe}\orcidlink{0000-0002-3972-1978}
Alyssa Bulatek,\affref{uflorida}\orcidlink{0000-0002-4407-885X} 
Claire Cook,\affref{kansas}
Jonathan D. Henshaw,\affrefTwo{ljmu}{mpia}\orcidlink{0000-0001-9656-7682}
\newauthor
Katharina Immer,\affref{eso}\orcidlink{0000-0003-4140-5138}
Namitha Issac,\affref{shao}\orcidlink{0000-0002-7881-689X}
Desmond Jeff,\affref{uflorida}\orcidlink{0000-0003-0416-4830} 
Fu-Heng Liang,\affrefTwo{ari_heidelberg}{eso}\orcidlink{0000-0003-2496-1247}
Steven N. Longmore,\affrefTwo{ljmu}{COOL}\orcidlink{0000-0001-6353-0170}
\newauthor
Elisabeth A.C. Mills,\affref{kansas}\orcidlink{0000-0001-8782-1992}
Sergio Martín,\affrefTwo{eso_chile}{jao}\orcidlink{0000-0001-9281-2919}
Xing Pan,\affrefThree{nanjing}{nanjing_key}{cfa}\orcidlink{0000-0003-1337-9059}
Qizhou Zhang,\affref{cfa}\orcidlink{0000-0003-2384-6589}
John Bally,\affref{colorado}\orcidlink{0000-0001-8135-6612}
\newauthor
Cara Battersby,\affref{uconn}\orcidlink{0000-0002-6073-9320}
Laura Colzi,\affref{cab_csic}\orcidlink{0000-0001-8064-6394}
Paul T. P. Ho, \affref{iaa_taipei}\orcidlink{0000-0002-3412-4306}
Izaskun Jim\'enez-Serra,\affref{cab_csic}\orcidlink{0000-0003-4493-8714}
\newauthor
J.~M.~Diederik Kruijssen,\affref{COOL}\orcidlink{0000-0002-8804-0212}
Maya A.~Petkova,\affref{chalmers}\orcidlink{0000-0002-6362-8159}
Mattia C. Sormani,\affref{clap}\orcidlink{0000-0001-6113-6241}
Robin G. Tress,\affref{iop_epfl}\orcidlink{0000-0002-9483-7164}
\newauthor
Jennifer Wallace,\affref{uconn}\orcidlink{0009-0002-7459-4174}
J. Armijos-Abenda\~no,\affref{oaq}\orcidlink{0000-0003-3341-6144}
Lucia Armillotta,\affref{inaf_arcetri}\orcidlink{0000-0002-5708-1927}
N. Bijas,\affref{jbca}\orcidlink{0000-0002-6398-7530}
\newauthor
Rojita Buddhacharya,\affref{ljmu}\orcidlink{0009-0004-0685-7678}
Laura A. Busch,\affref{mpe}
Natalie O. Butterfield,\affref{nrao}\orcidlink{0000-0002-4013-6469}
M\'elanie Chevance\affrefTwo{ita_heidelberg}{COOL}\orcidlink{0000-0002-5635-5180}
\newauthor
Ana Karla D\'iaz-Rodr\'iguez,\affref{ukarcnode}\orcidlink{0000-0001-9112-6474}
Christoph Federrath,\affref{anu}\orcidlink{0000-0002-0706-2306}
Rub\'{e}n Fedriani,\affref{iaa_csic}\orcidlink{0000-0003-4040-4934}
Pablo Garc{\'i}a,\affrefTwo{ucn}{cassaca}\orcidlink{0000-0002-8586-6721}
\newauthor
Qi-Lao Gu,\affref{shao}
H Perry Hatchfield,\affref{uconn}\orcidlink{0000-0003-0946-4365} 
Rebecca J. Houghton,\affref{ljmu}\orcidlink{0000-0002-9723-1088}
Yue Hu,\affref{ias}\orcidlink{0000-0002-8455-0805}
Janik Karoly,\affref{ucl}\orcidlink{0000-0001-5996-3600}
\newauthor
Ralf S.\ Klessen,\affrefThree{ita_heidelberg}{izw_heidelberg}{cfa}\thanks{Elizabeth S.\ and Richard M.\ Cashin Fellow at the Radcliffe Institute for Advanced Studies at Harvard University, 10 Garden Street, Cambridge, MA 02138, USA}\orcidlink{0000-0002-0560-3172}
Mark R. Krumholz,\affref{anu}\orcidlink{0000-0003-3893-854X}
Xunchuan Liu, \affref{shao}\orcidlink{0000-0001-8315-4248}
Farideh Mazoochi,\affref{ipm}\orcidlink{0000-0003-3390-4893}
\newauthor
Francisco Nogueras-Lara,\affrefTwo{iaa_csic}{eso}\orcidlink{0000-0002-6379-7593}
Dylan Par\'e,\affrefTwo{jao}{nrao}\orcidlink{0000-0002-5811-0136}
Denise Riquelme-V\'asquez,\affref{ulaserena}\orcidlink{0000-0001-5389-0535}
V\'ictor M. Rivilla,\affref{cab_csic}\orcidlink{0000-0002-2887-5859}
\newauthor
Miriam G. Santa-Maria,\affrefTwo{uflorida}{iff_csic}\orcidlink{0000-0002-3941-0360}
Anika Schmiedeke,\affref{gbo}\orcidlink{0000-0002-1730-8832}
Yoshiaki Sofue,\affref{utokyo}\orcidlink{0000-0002-4268-6499}
Volker Tolls,\affref{cfa}\orcidlink{0000-0003-1841-2241}
\newauthor
Q. Daniel Wang,\affref{umass}\orcidlink{0000-0002-9279-4041}
Gwenllian M. Williams,\affref{aberystwyth}\orcidlink{0000-0001-5933-2147}
Fengwei Xu,\affrefTwo{kiaa_pku}{pku_astro}\orcidlink{0000-0001-5950-1932},
and Suinan Zhang\affrefTwo{ist}{naoj}\orcidlink{0000-0002-8389-6695}
\\
$^{*}$Author affiliations are listed at the end of the paper
}

\date{Accepted 2026 February 02. Received 2026 February 02; in original form 2025 July 18}

\pubyear{2026}

\begin{document}
\label{firstpage}
\pagerange{\pageref{firstpage}--\pageref{lastpage}}
\maketitle

\begin{abstract}
We release the intermediate-width spectral window data from the ALMA Central Molecular Zone Exploration Survey (ACES) large program, which covers SiO (2--1), SO (2$_2$--1$_1$), \htcop{} (1--0), \htcn{} (1--0), \hntc{} (1--0), and \hcfn{} (1--0), among other molecular line transitions, with an angular resolution of $\sim$2\arcsec{} and a velocity resolution of 1.7~\kms{}. The full cubes of the two spectral windows as well as the key data products will be available to the community. We also present the integrated brightness, peak brightness, centroid velocity, and Galactic longitude-velocity maps of the six lines. We briefly discuss morphological correlations between the continuum and the molecular line emission, and brightness ratios between pairs of isotopologue or isotopomer lines. We highlight features and trends in the data that will be followed up in upcoming ACES science papers.
\end{abstract}

\begin{keywords}
Galactic: centre --- ISM: structure
\end{keywords}

\section{Introduction}
\label{sec:intro}

The Central Molecular Zone (CMZ) usually refers to the central $\sim$500~pc of our Galaxy, where 2--6$\times$ 10$^7$~\msol{} of molecular gas is detected \citep{Ferriere2007,Henshaw2023}. The distance to the supermassive black hole Sgr~A* inside the CMZ is measured to 8.277~kpc \citep{Gravity2022}, which is often adopted as the distance to the CMZ. The measured physical and chemical properties of the molecular gas are extreme compared to those of clouds in the Galactic disk, with e.g., higher temperatures \citep[$\gtrsim$50--100~K vs $\sim$10--20~K;][]{Ao2013,Mills2013b,Ginsburg2016,Immer2016,Krieger2017,Lu2017}, stronger non-thermal motions \citep[FWHM$\sim$5--10~\kms{} vs $\lesssim$1~\kms{} at 0.1~pc to pc scales;][]{Liu2013,Rathborne2014a,Henshaw2016b}, and stronger magnetic fields \citep[$\sim$mG vs $\sim$0.01--0.1~mG;][]{Pillai2015,Mangilli2019,Guan2021,Lu2024}. The star formation rate (SFR) in the CMZ has been measured to be $\sim$0.07~\msol{}\,yr$^{-1}$ over the last 5~Myr, which is about 10\% of the expected value given the well-established linear correlation between the amount of dense molecular gas and the SFR \citep{Kauffmann2017b,Barnes2017,Lu2019a,Henshaw2023}. Understanding how the physical conditions of the molecular gas are related to the inefficient star formation in the CMZ is an ongoing challenge that requires the combined insight from both observations and numerical simulations to elucidate the key physics \citep{Kruijssen2014a, Kruijssen19, krumholz15, Krumholz16,Tress2020,Sormani2022}.

A critical observational part of this challenge is to characterize the properties of the molecular gas with CMZ-wide surveys of molecular lines. There exist several surveys using single dish telescopes targeting the CO lines \citep[e.g.,][]{Bally1987,Oka1998a,Oka2007,Eden2020}, the NH$_3$ inversion lines \citep[e.g.,][]{Huettemeister1993,Arai2016,Longmore2017}, and particularly, several 3~mm lines including SiO, \htcn{}, and \htcop{} that trace shocks or dense gas \citep[e.g.,][]{Martin-Pintado1997,Requelme2010,Jones2012}. Such observations have provided important constraints to modeling gas dynamics \citep[e.g.,][]{Kruijssen2015,Henshaw2016a}, heating and cooling of the gas \citep[e.g.,][]{Ao2013,Ginsburg2016,Immer2016}, and chemistry \citep[e.g.,][]{Martin-Pintado1997,Martin-Pintado2000,Requena-Torres2006,Requena-Torres2008,Martin2008,Jones2012}. However, interferometric observations with resolutions $\lesssim$2\farcs{5} are necessary to resolve molecular gas down to the spatial scales of individual dense cores ($\sim$0.1~pc) and uncover the immediate environment of star formation.

The ALMA CMZ Exploration Survey (ACES) Large Program uniformly covers the central $\sim$200~pc of the CMZ using the Atacama Large Millimeter/submillimeter Array (ALMA) with its 3~mm band (Band~3), generating the largest mosaic ALMA has yet produced ($>$1000 arcmin$^2$) with an unprecedented resolution of $\sim$2\arcsec{} \citep[Project code: 2021.1.00172.L, PI: S.\ Longmore;][hereafter \citetalias{Longmore2026}]{Longmore2026}. The main scientific objectives are to derive the physical, kinematic and chemical properties of the dense gas and protostellar and prestellar core population in the inner CMZ and use this to build a global understanding of star formation, feedback, and the mass flows and energy cycles across the Central Molecular Zone.

The observations captured the 3~mm continuum with an effective bandwidth of 4.6~GHz, which is released in \citet[][hereafter \citetalias{Ginsburg2026}]{Ginsburg2026}. In addition to the continuum emission, three different setups of spectral windows (SPWs) were used to sample spectral lines: two narrow SPWs each with a total bandwidth of 58.59~MHz ($\sim$200~\kms{}) and an effective resolution of 0.06~MHz ($\sim$0.2~\kms{}), to cover HNCO (4--3) and \hcop{} (1--0), respectively, which are released in \citet[][hereafter \citetalias{Walker2026}]{Walker2026}; two wide SPWs each with a total bandwidth of 1.875~GHz and an effective resolution of 0.976~MHz ($\sim$2.9~\kms{}), to cover various lines including CS (2--1), which are released in \citet[][hereafter \citetalias{Hsieh2026}]{Hsieh2026}; and two intermediate-width SPWs each with a total bandwidth of 0.469~GHz and a resolution of 0.488~MHz ($\sim$1.7~\kms{}, which characterises the ability to distinguish spectral features and is twice the physical channel width\footnote{See \url{https://help.almascience.org/kb/articles/new-default-for-the-bandwidth-used-for-sensitivity-in-the-ot} for details.}), which are released in this paper. The 12-m array, 7-m array, and the Total Power (TP, i.e., single-dish) array of ALMA were used for recovering structures of different spatial scales. The bandwidths of the spectral windows of the three arrays are different, with those of the 7-m and TP arrays slightly broader than that of the 12-m array to ensure effective data combination with the latter. In the following, we refer to the two intermediate-width SPWs as 12-m SPWs 25 and 27 (see \autoref{tab:lines}), respectively.

\begin{figure*}
    \centering
    \includegraphics[width=1\textwidth]{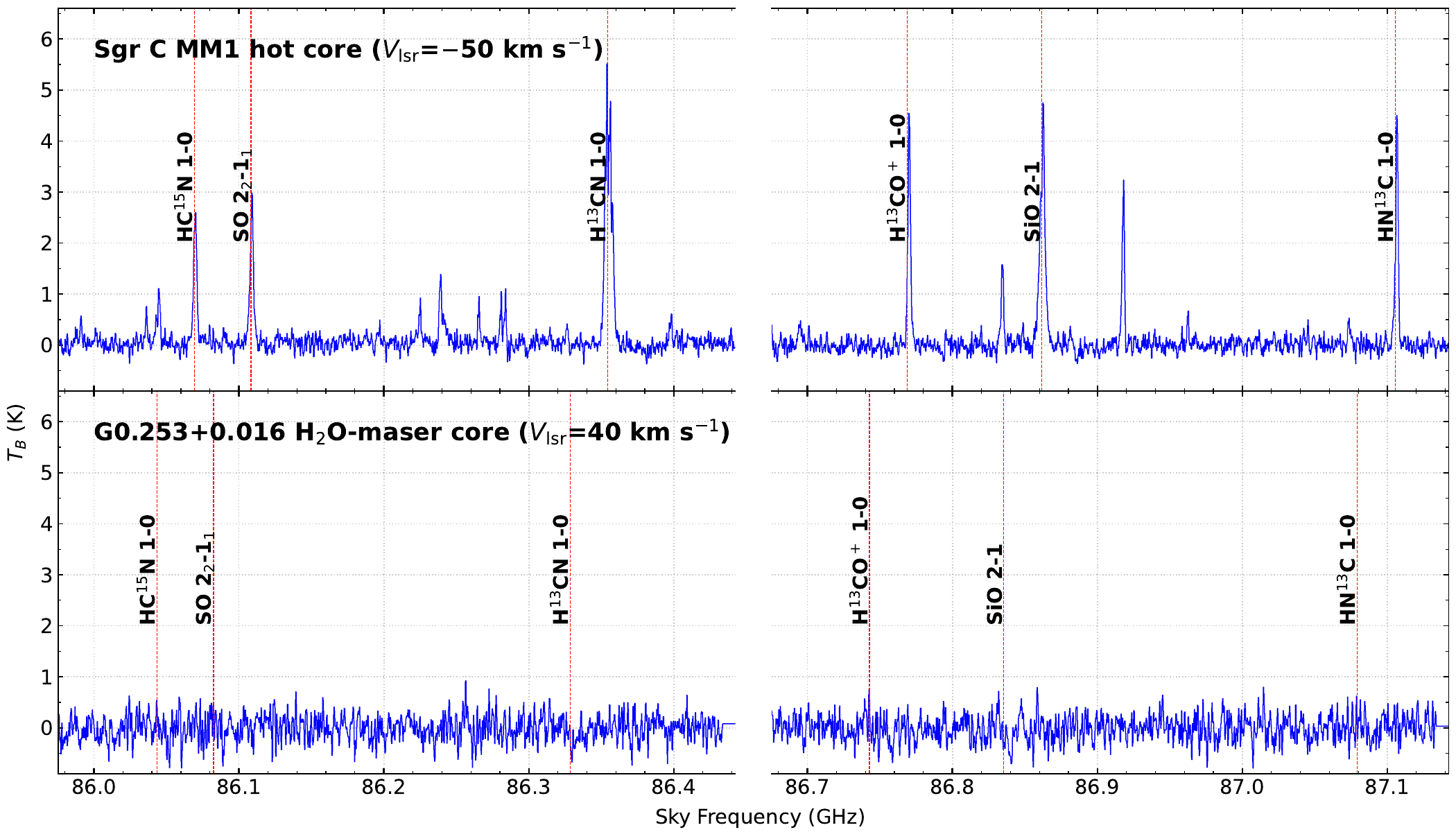}
    \caption{Typical spectra of the two SPWs toward a hot core (top) and a relatively quiescent core (bottom), respectively. The spectra are extracted and averaged from a circle with a radius of 0\farcs{75}, centered at ($l$, $b$)=(359.436\arcdeg{}, $-$0.104\arcdeg{}) and (0.261\arcdeg{}, 0.016\arcdeg{}), respectively. The six transitions presented in this paper are labeled. There are other line emission features in the two SPWs that will not be discussed in this data release paper.
    }
    \label{fig:spec}
\end{figure*}

\section{Observations}
\label{sec:obs}

General information about the observations can be found in the survey description paper \citepalias{Longmore2026} and other data release papers \citepalias{Ginsburg2026,Walker2026,Hsieh2026}. Here we focus on the 12-m SPWs 25 and 27. The 12-m SPW 25 was centered at a rest frequency of 86.2~GHz and covered frequencies between 85.96~GHz and 86.43~GHz, which allows it to sample several spectral lines including \hcfn{} (1--0), SO (2$_2$--1$_1$), and \htcn{} (1--0). The 12-m SPW 27 was centered at a rest frequency of 86.9~GHz, covering frequencies between 86.66~GHz and 87.13~GHz and spectral lines including \htcop{} (1--0), SiO (2--1), and \hntc{} (1--0). \autoref{fig:spec} presents ACES spectra from a hot core in the top panel and those from a relatively quiescent core in the bottom. The difference in their chemistry is evident. Key properties of the six strong lines are listed in \autoref{tab:lines}. In the following, we focus on those six lines and present their images and statistics, while other lines will be included in the data release but will not be discussed here.

\begin{table*}
\caption{Spectral configuration of the 12-m SPWs 25 and 27, and key spectral lines covered. The upper and lower frequency bounds of the SPWs are those of the 12-m array data.\label{tab:lines}}
\centering
\begin{tabular}{ccccccccc}
\hline
SPW id & $\nu_\text{L}$ & $\nu_\text{U}$ & Molecule & Transition & Rest frequency & $E_\text{u}/k$ & $\log$($A_{\rm ul}$) & $n_{\rm crit}$$^a$ \\
(12-m/7-m/TP) & (MHz) & (MHz) & & & (MHz) & (K) & (s$^{-1}$) & (cm$^{-3}$) \\
\hline
25/16/17 & 85965.6 & 86434.4 & HC$^{15}$N & $J$=1--0 & 86054.97 & 4.13 & $-$4.6569 & $2\times10^6$ \\
& & & SO & $N_J$=2$_2$--1$_1$ & 86093.95 & 19.3 & $-$5.2798 & $2\times10^5$ \\
& & & \htcn{} & $J$=1--0    & 86339.92 & 4.14 & 
$-$4.6526 & $2\times10^6$ \\
27/18/19 & 86665.6 & 87134.4 & \htcop{} & $J$=1--0   & 86754.30 & 4.16 & $-$4.4141 & $2\times10^5$ \\
& & & SiO & $J$=2--1        & 86846.96 & 6.25 & $-$4.5336 & $2\times10^5$ \\
& & & \hntc{} & $J$=1--0    & 87090.85 & 4.18 & $-$4.5702 & $4\times10^5$ \\
\hline
\end{tabular}
\\
$^a$ Critical densities are calculated as $n_{\rm crit} = A_{\rm ul}/\gamma_{\rm ul}$, where $A_{\rm ul}$ is the Einstein coefficient and $\gamma_{\rm ul}$ is the collisional rate, whose values at 100~K are taken from the Leiden Atomic and Molecular Database \citep{Schoier2005}.
\end{table*}

\section{Data processing}
\label{sec:data}

\subsection{Data reduction}
\label{subsec:data_reduction}

The data were calibrated and imaged with the CASA package 6.5.4 \citep{CASA2022}. Procedures of continuum subtraction and array combination are described in \citetalias{Walker2026}. Note that in this data release, we have adopted the ``feather-only'' approach (see \citetalias{Walker2026}) to combine images from the three different arrays. In Appendix~\ref{app_sec:comb_miriad}, we describe an alternative method that employs the software package MIRIAD\footnote{\url{https://www.astro.umd.edu/~teuben/miriad/}} \citep{Sault1995} and a different approach to combine interferometer and single dish data. We have chosen not to apply this alternative method to the full ACES dataset, but invite interested readers to refer to Appendix~\ref{app_sec:comb_miriad} and references therein for further details.

The images of the two SPWs were produced with the \textit{tclean} task in CASA. The native channel width is 0.244~MHz ($\sim$ 0.85~\kms{}), half of the spectral resolution, which was preserved in the imaging processes. For different regions in the large mosaic, slightly different imaging parameters (e.g., robustness number, pixel size) were chosen. Image cubes of different regions were then smoothed to a common beam and were stitched together to produce the final, full-CMZ mosaics. The cube of 12-m SPW 25 has a beam size of 3.17\arcsec{} $\times$ 2.15\arcsec{} with a position angle of $-$85.9\arcdeg{}, while the cube of 12-m SPW 27 has a slighter larger beam of 3.25\arcsec{} $\times$ 2.26\arcsec{} with a position angle of $-$85.7\arcdeg{}. The geometric mean of the beams is $\sim$2.6\arcsec{}, corresponding to a linear resolution of 0.1~pc at a distance of 8.277~kpc. The image rms noise per 0.85~\kms{} channel is not uniform across the mosaics, but varies between regions typically in the range of 3--5~\mjypbm{}, and can be much higher in the area surrounding regions of strong emission that limit the dynamic range.

\subsection{Data products}
\label{subsec:data_products}

\begin{figure*}
\centering
\includegraphics[width=1.0\textwidth]{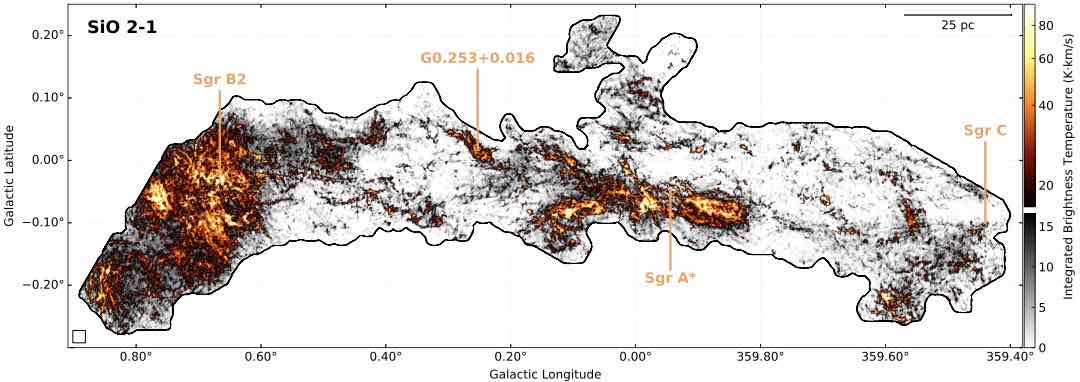}
\includegraphics[width=1.0\textwidth]{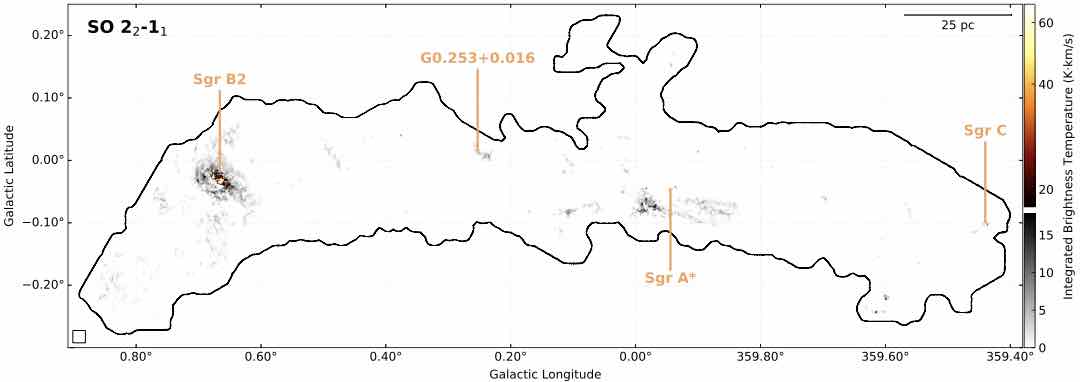}
\includegraphics[width=1.0\textwidth]{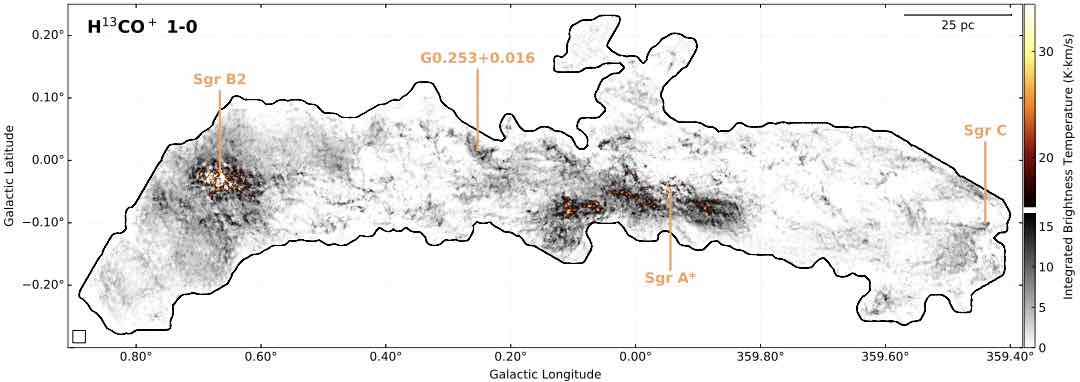}
\caption{Integrated brightness temperature maps of SiO (2--1), SO (2$_2$--1$_1$), and \htcop{} (1--0). The two-piece colour bars are a combination of a linear scale between 0 and approximately 5$\sigma$ and a logarithmic scale above approximately 5$\sigma$ up to the 99.9 percentile of the data.}
\label{fig:mom0}
\end{figure*}

\begin{figure*}
\centering
\includegraphics[width=1.0\textwidth]{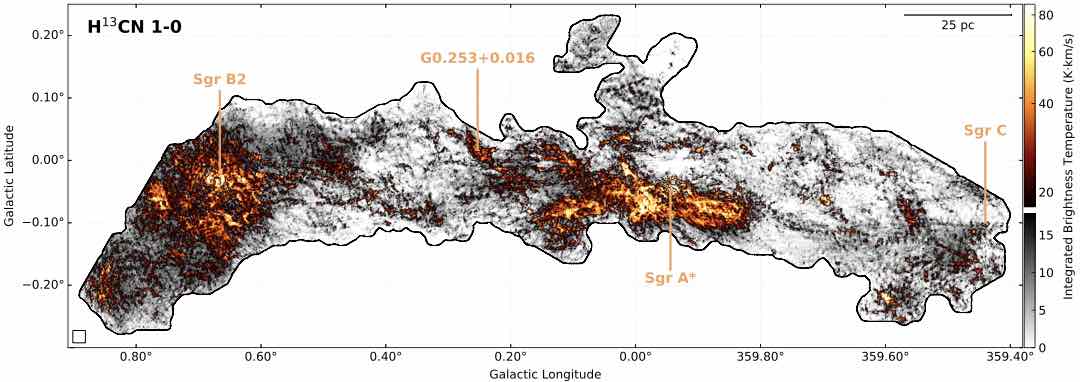}
\includegraphics[width=1.0\textwidth]{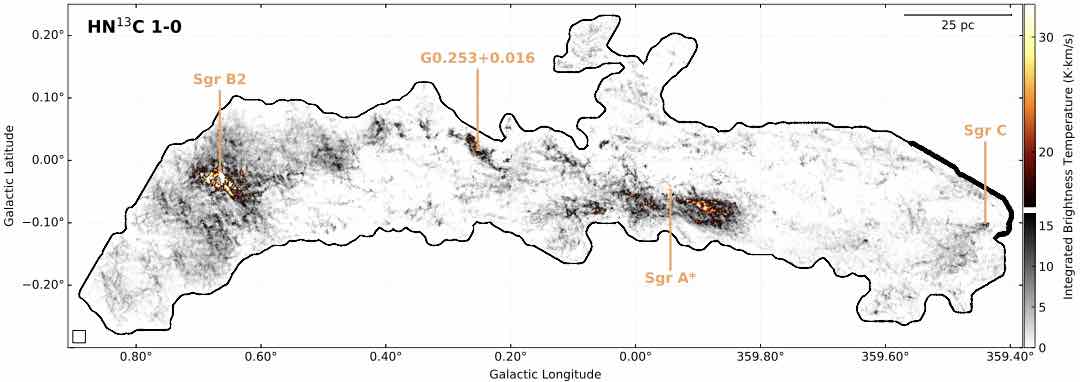}
\includegraphics[width=1.0\textwidth]{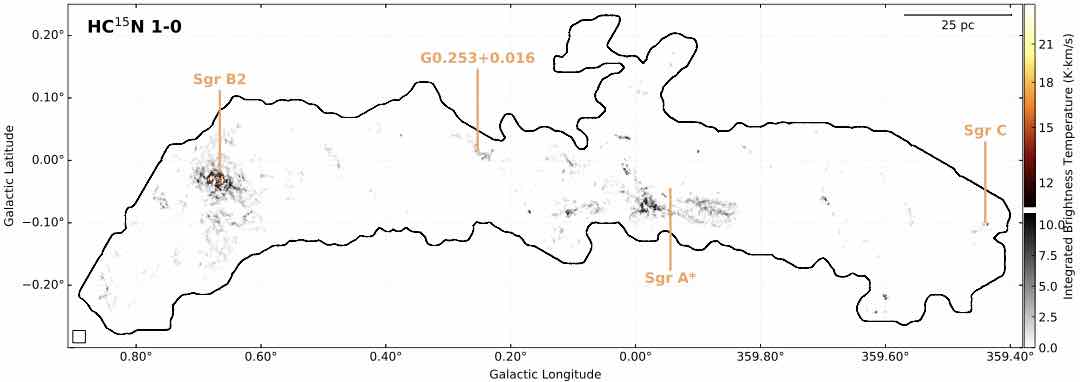}
\caption{Integrated brightness temperature maps of \htcn{} (1--0), \hntc{} (1--0), and \hcfn{} (1--0). The two-piece colour bars are a combination of a linear scale between 0 and approximately 5$\sigma$ and a logarithmic scale above approximately 5$\sigma$ up to the 99.9 percentile of the data.}
\label{fig:mom0-continued}
\end{figure*}

\begin{figure*}
\centering
\includegraphics[width=1.0\textwidth]{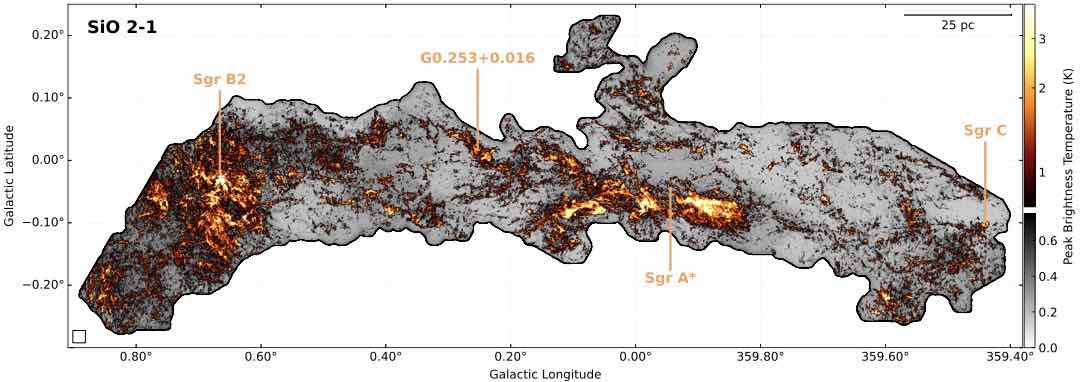}
\includegraphics[width=1.0\textwidth]{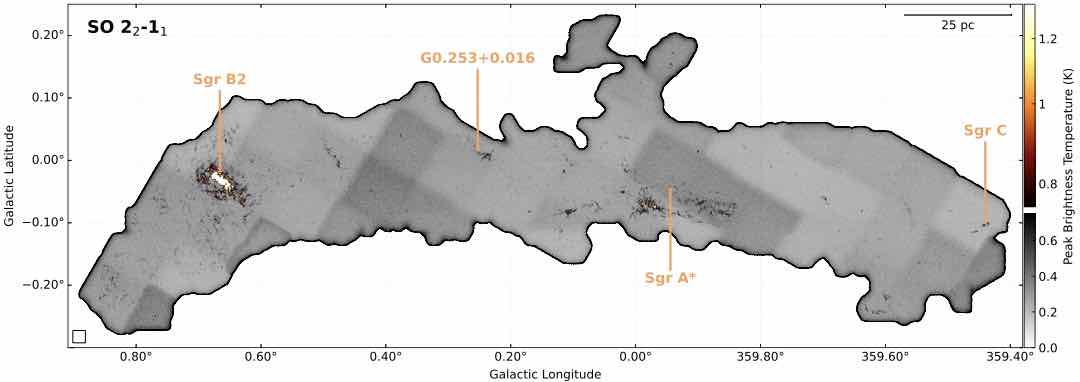}
\includegraphics[width=1.0\textwidth]{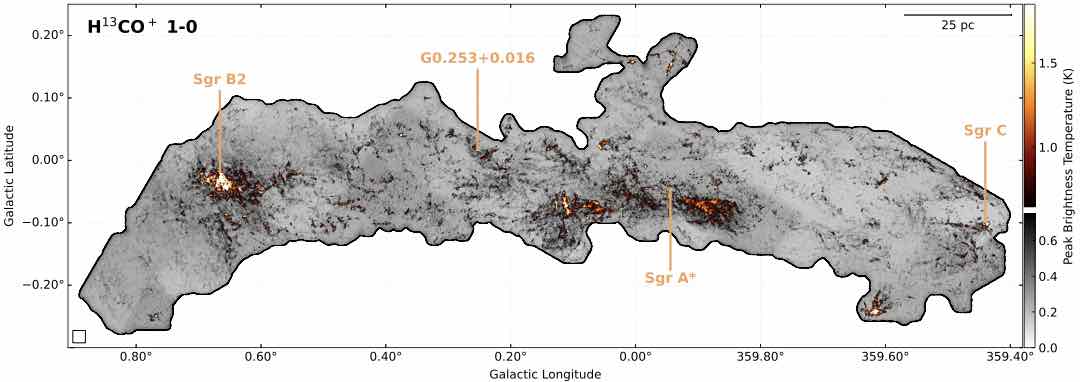}
\caption{Peak brightness temperature maps of SiO (2--1), SO (2$_2$--1$_1$), and \htcop{} (1--0). The two-piece colour bars are a combination of a linear scale between 0 and approximately 5$\sigma$ and a logarithmic scale above approximately 5$\sigma$ up to the 99.9 percentile of the data.}
\label{fig:mom8}
\end{figure*}

\begin{figure*}
\centering
\includegraphics[width=1.0\textwidth]{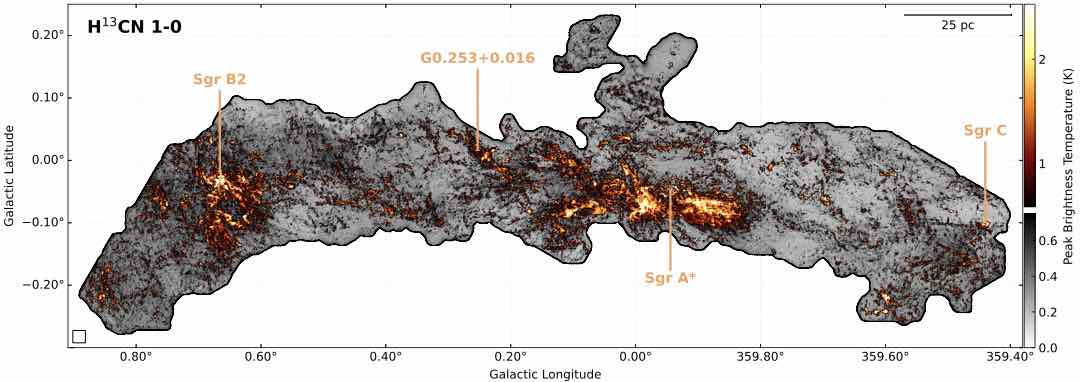}
\includegraphics[width=1.0\textwidth]{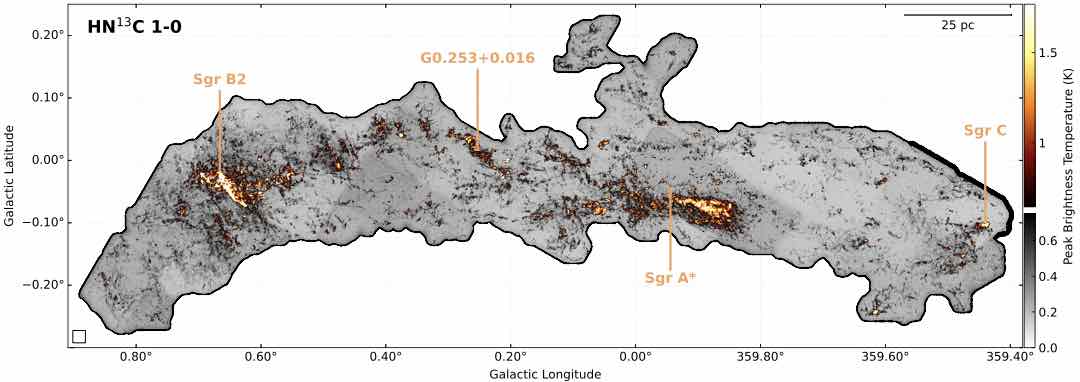}
\includegraphics[width=1.0\textwidth]{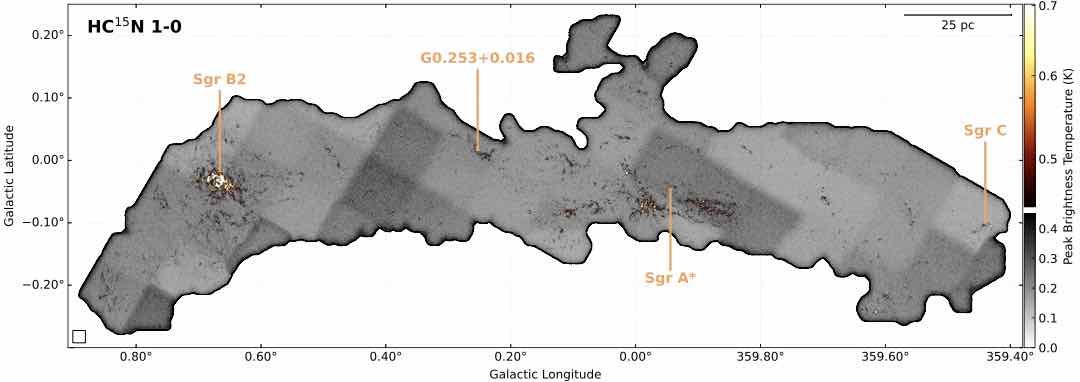}
\caption{Peak brightness temperature maps of \htcn{} (1--0), \hntc{} (1--0), and \hcfn{} (1--0). The two-piece colour bars are a combination of a linear scale between 0 and approximately 5$\sigma$ and a logarithmic scale above approximately 5$\sigma$ up to the 99.9 percentile of the data.}
\label{fig:mom8-continued}
\end{figure*}

\begin{figure*}
\centering
\includegraphics[width=1.0\textwidth]{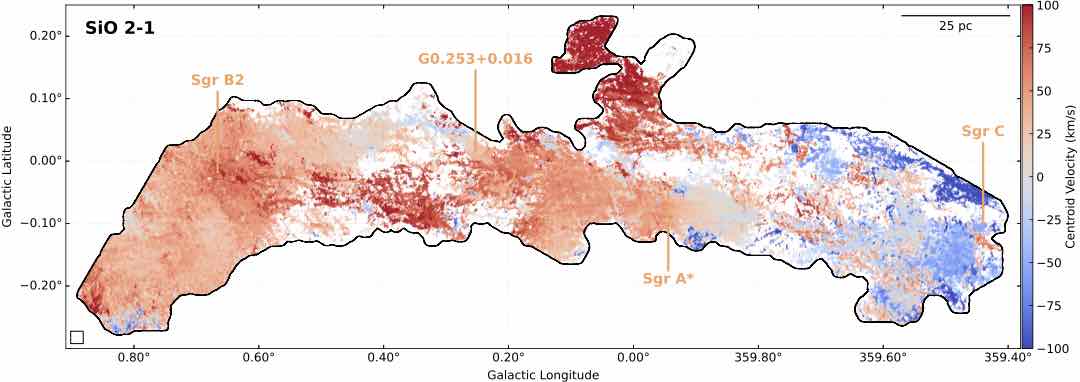}
\includegraphics[width=1.0\textwidth]{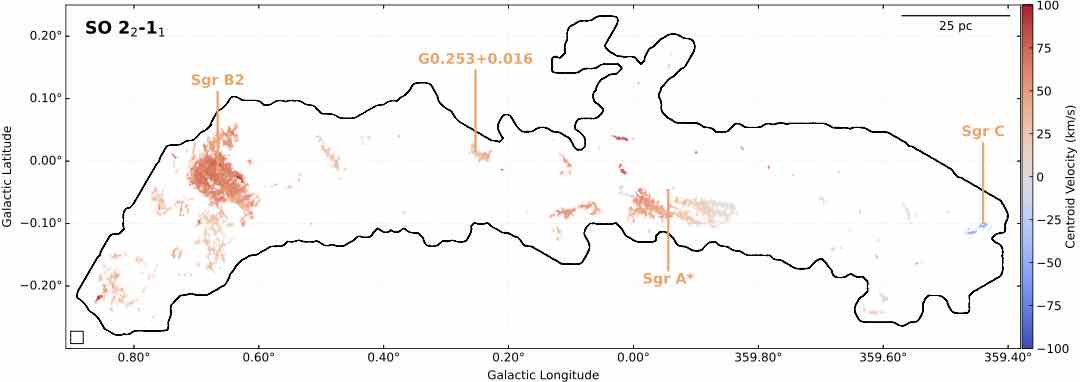}
\includegraphics[width=1.0\textwidth]{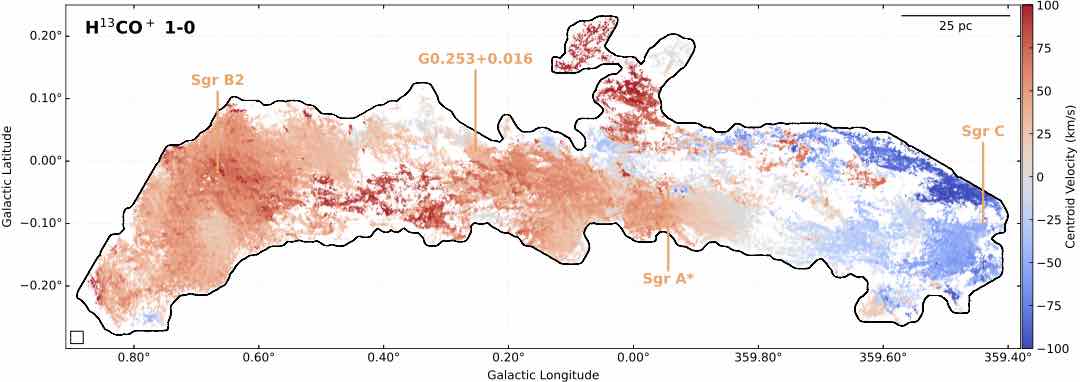}
\caption{Centroid velocity maps of SiO (2--1), SO (2$_2$--1$_1$), and \htcop{} (1--0). All the colour bars are linear and truncated between $-$100 and 100~\kms{} to highlight the velocity structures.}
\label{fig:mom1}
\end{figure*}

\begin{figure*}
\centering
\includegraphics[width=1.0\textwidth]{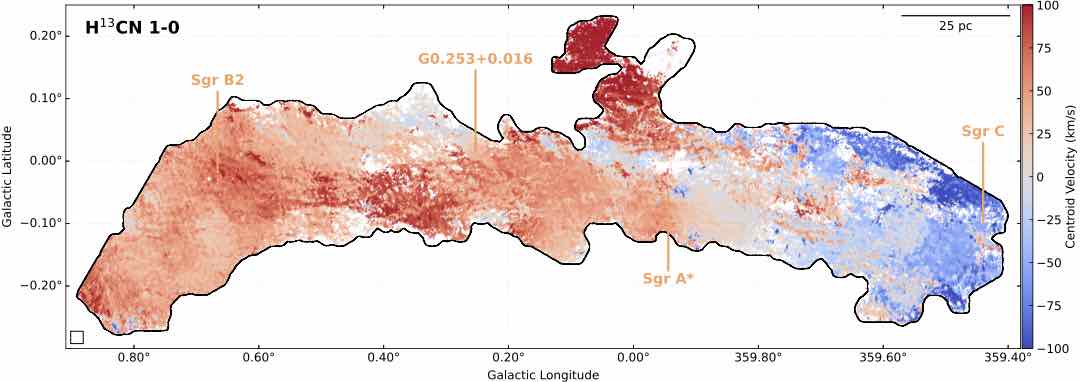}
\includegraphics[width=1.0\textwidth]{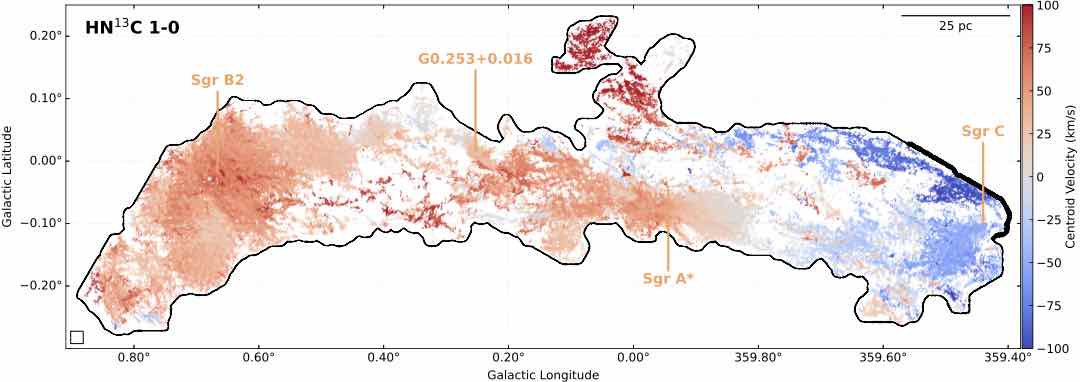}
\includegraphics[width=1.0\textwidth]{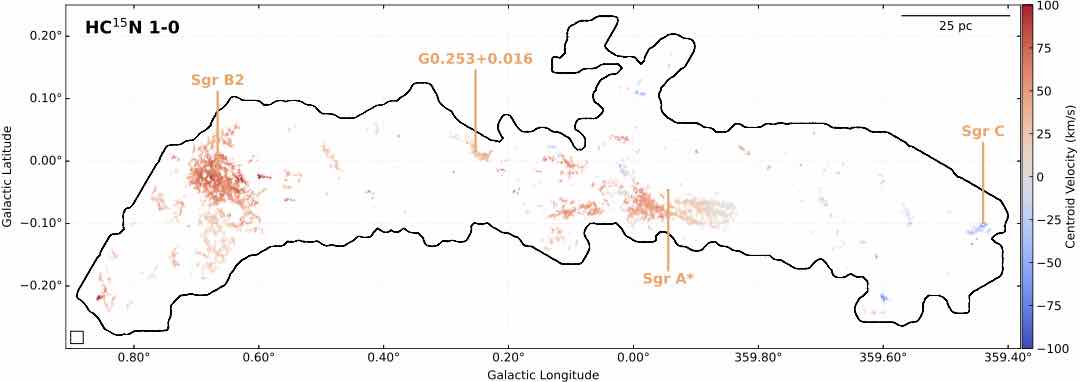}
\caption{Centroid velocity maps of \htcn{} (1--0), \hntc{} (1--0), and \hcfn{} (1--0). All the colour bars are linear and truncated between $-$100 and 100~\kms{} to highlight the velocity structures.}
\label{fig:mom1-continued}
\end{figure*}

\begin{figure*}
\centering
\includegraphics[width=1.0\textwidth]{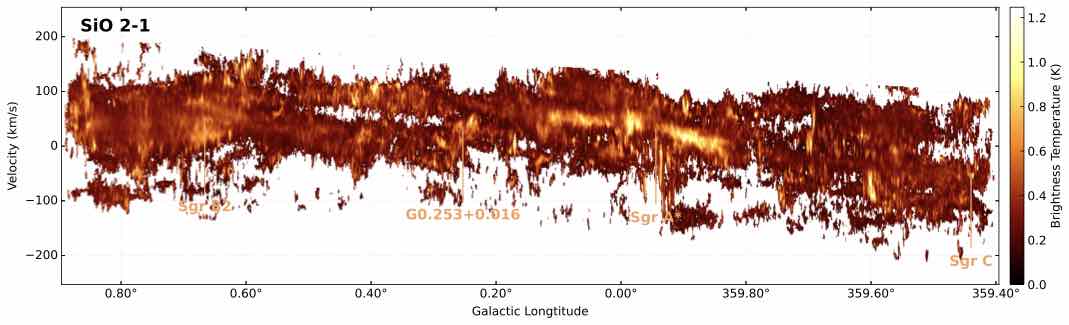}
\includegraphics[width=1.0\textwidth]{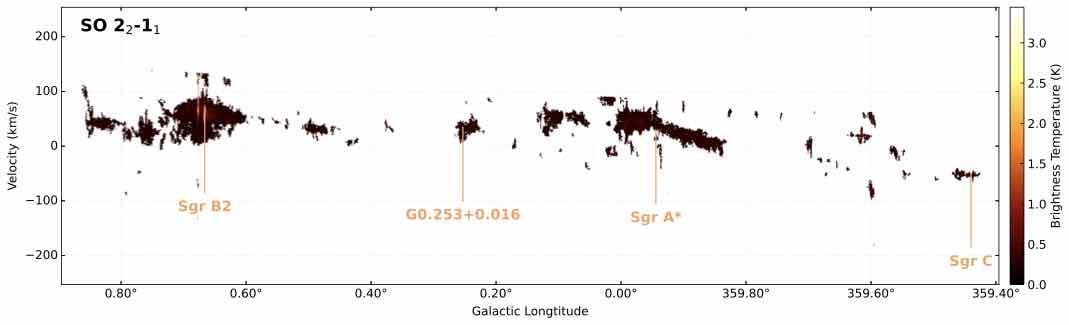}
\includegraphics[width=1.0\textwidth]{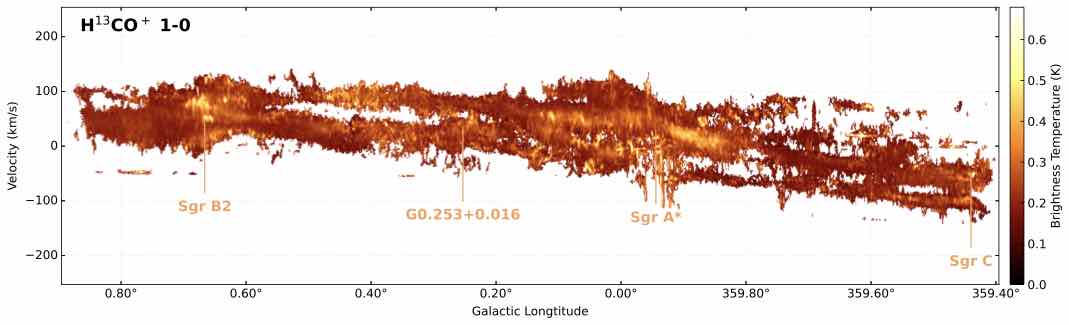}
\caption{Position-velocity diagrams of SiO (2--1), SO (2$_2$--1$_1$), and \htcop{} (1--0). Here the intensities are the mean values of the cut along the Galactic latitude. The velocities are truncated by the $l$-\vlsr{} mask specified in Section~\ref{subsec:data_products}.
}
\label{fig:pv_mean}
\end{figure*}

\begin{figure*}
\centering
\includegraphics[width=1.0\textwidth]{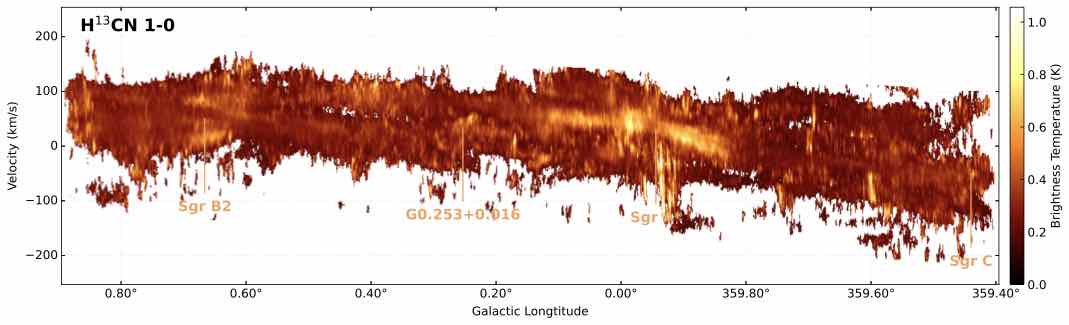}
\includegraphics[width=1.0\textwidth]{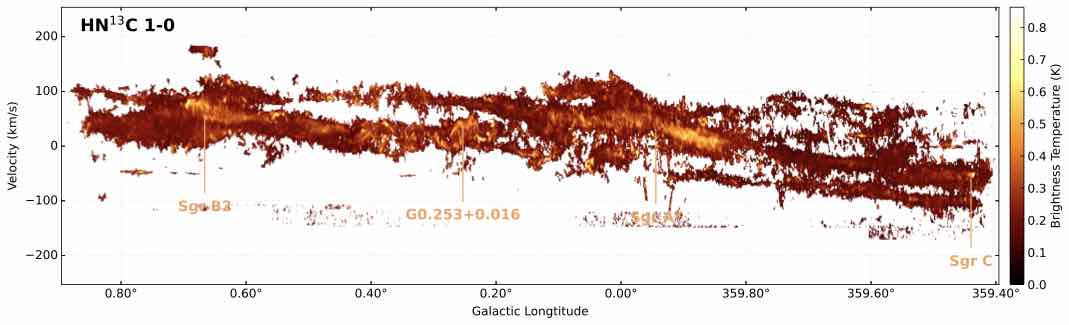}
\includegraphics[width=1.0\textwidth]{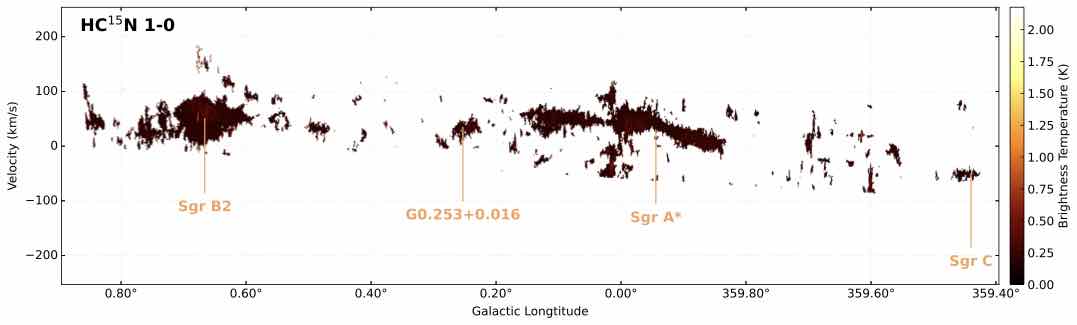}
\caption{Position-velocity diagrams of \htcn{} (1--0), \hntc{} (1--0), and \hcfn{} (1--0). Here the intensities are the mean values of the cut along the Galactic latitude. The velocities are truncated by the $l$-\vlsr{} mask specified in Section~\ref{subsec:data_products}.
}
\label{fig:pv_mean-continued}
\end{figure*}

\begin{figure*}
\centering
\includegraphics[width=1.0\textwidth]{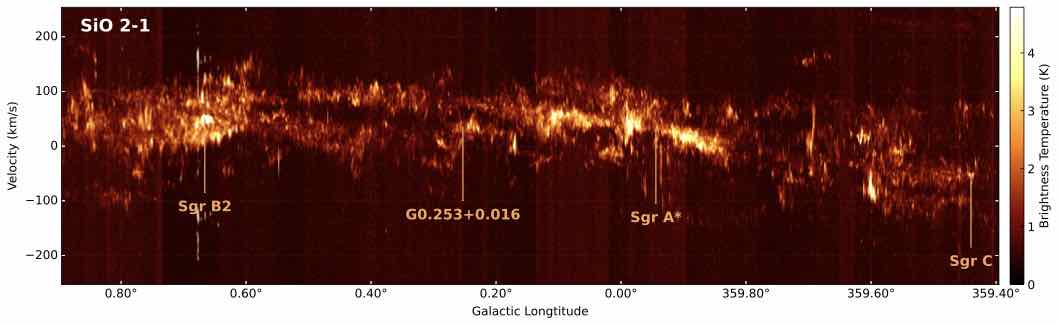}
\includegraphics[width=1.0\textwidth]{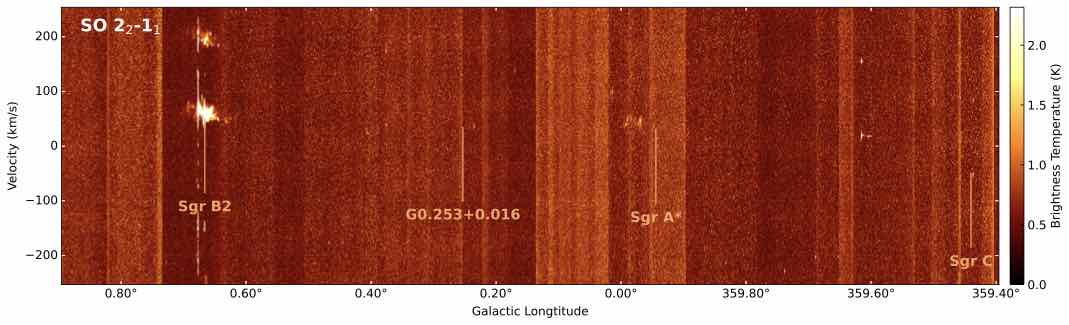}
\includegraphics[width=1.0\textwidth]{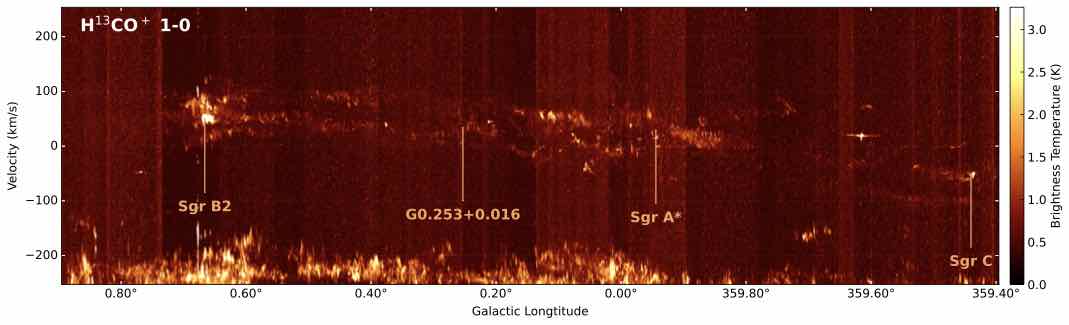}
\caption{Position-velocity diagrams of SiO (2--1), SO (2$_2$--1$_1$), and \htcop{} (1--0). Here the intensities are the maximum values of the cut along the Galactic latitude, without the $l$-\vlsr{} mask.
}
\label{fig:pv_peak}
\end{figure*}

\begin{figure*}
\centering
\includegraphics[width=1.0\textwidth]{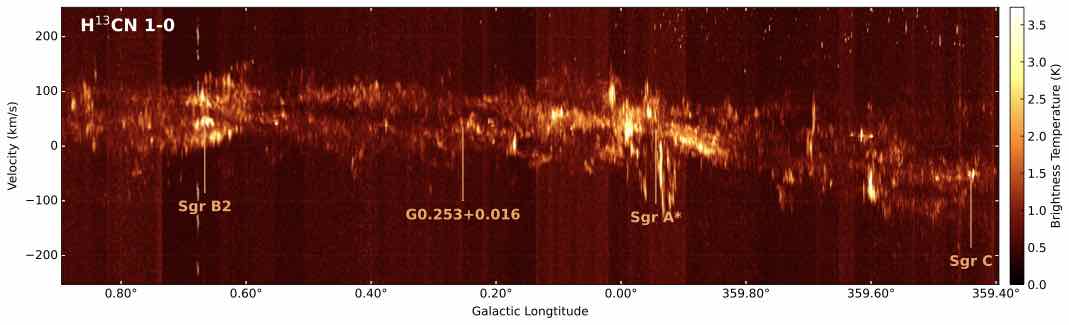}
\includegraphics[width=1.0\textwidth]{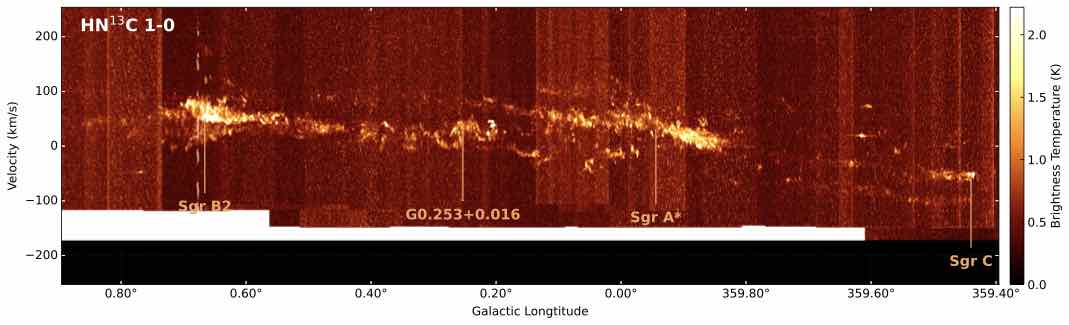}
\includegraphics[width=1.0\textwidth]{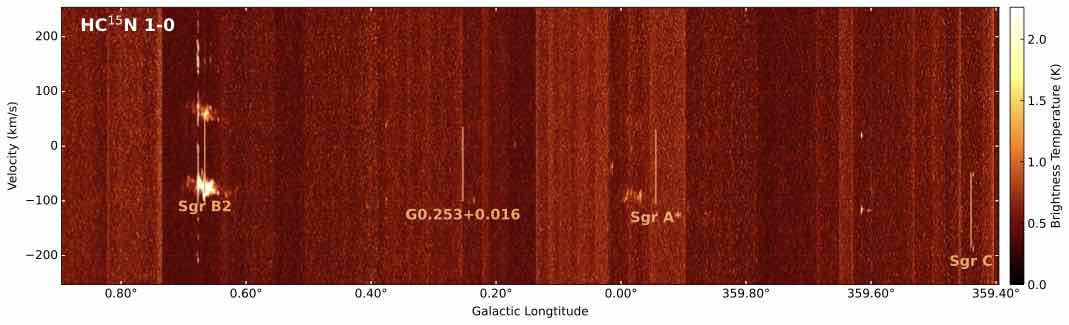}
\caption{Position-velocity diagrams of \htcn{} (1--0), \hntc{} (1--0), and \hcfn{} (1--0). Here the intensities are the maximum values of the cut along the Galactic latitude, without the $l$-\vlsr{} mask.}
\label{fig:pv_peak-continued}
\end{figure*}

The full cubes of the two SPWs are released at \url{https://almascience.org/alma-data/lp/aces}. For the six lines highlighted in \autoref{fig:spec} and \autoref{tab:lines}, we additionally provide their integrated brightness temperature maps, centroid velocity maps, peak brightness temperature maps, and Galactic longitude-velocity as well as Galactic latitude-velocity maps. Except otherwise noted, a customized mask in the Galactic longitude-velocity plain ($l$-\vlsr{} mask hereafter) was applied when making the maps, which covers a velocity (\vlsr{}) range between $-$120 and 200~\kms{} at Galactic longitude of 0.9\arcdeg{} and between $-$220 and 100~\kms{} at Galactic longitude of $-$0.6\arcdeg{}, and linearly interpolates in between the two ends. These maps of the six lines are presented in Figures~\ref{fig:mom0}--\ref{fig:pv_peak-continued}.

There are several considerations for the data products that warrant emphasis:
\begin{itemize}
\item The rest frequency of a SiO rotational-vibrational transition ($J$=2--1, $v$=1), often showing maser emission, is 86243.37~MHz, which is close to the frequency of \htcn{}. This SiO maser is usually excited in the envelopes of AGB stars, and has been found to be populous in the CMZ \citep{Messineo2002,Messineo2020,Li2010,Borkar2020}. As a result, the \htcn{} data is contaminated by sporadic SiO masers. This is most evident in the Galactic longitude-velocity map in \autoref{fig:pv_peak} where the $l$-\vlsr{} mask is not applied. The other maps of \htcn{} presented in this paper have had the $l$-\vlsr{} mask applied to them, and therefore many, but not all, of the SiO masers have been excluded. 
\item The SO (2$_2$--1$_1$) and \hcfn{} (1--0) lines are close in rest frequency, and therefore contaminate each other in the Galactic longitude-velocity maps in \autoref{fig:pv_peak} where both lines can be clearly seen in the plot of the other line. To separate emission of the two lines, for SO (2$_2$--1$_1$), the maximum \vlsr{} boundaries of the $l$-\vlsr{} mask at $l=0.9$\arcdeg{} and $-0.6$\arcdeg{} are adjusted to 150 and 50~\kms{}, respectively, whereas for \hcfn{} (1--0), the minimum \vlsr{} boundaries of the $l$-\vlsr{} mask are 0 and $-100$~\kms{}, respectively. The contamination is alleviated but not fully removed. Users of the data are encouraged to design a more sophisticated mask to effectively separate the two lines.
\item The \htcop{} (1--0) and SiO (2--1) lines are also close in rest frequency. In \autoref{fig:pv_peak}, the SiO emission appears at the blue-shifted velocities ($<-200$~\kms{}) in the plot of \htcop{}. To get rid of the SiO contamination, the minimum \vlsr{} boundaries of the $l$-\vlsr{} mask are $-$50 and $-150$~\kms{}, respectively. Again, users of the data could design their own masks to more effectively remove the SiO contamination.
\item The \hntc{} (1--0) line is on the high-frequency edge of the 12-m SPW 27, and therefore its cube is truncated at the blue-shifted side, as shown in \autoref{fig:pv_peak}.
\end{itemize}

\section{Discussion}
\label{sec:discussion}

\subsection{Morphological correlations between the continuum and the molecular emission}\label{subsec:dis_2dcorr}

\begin{figure}
\centering
\includegraphics[width=0.5\textwidth]{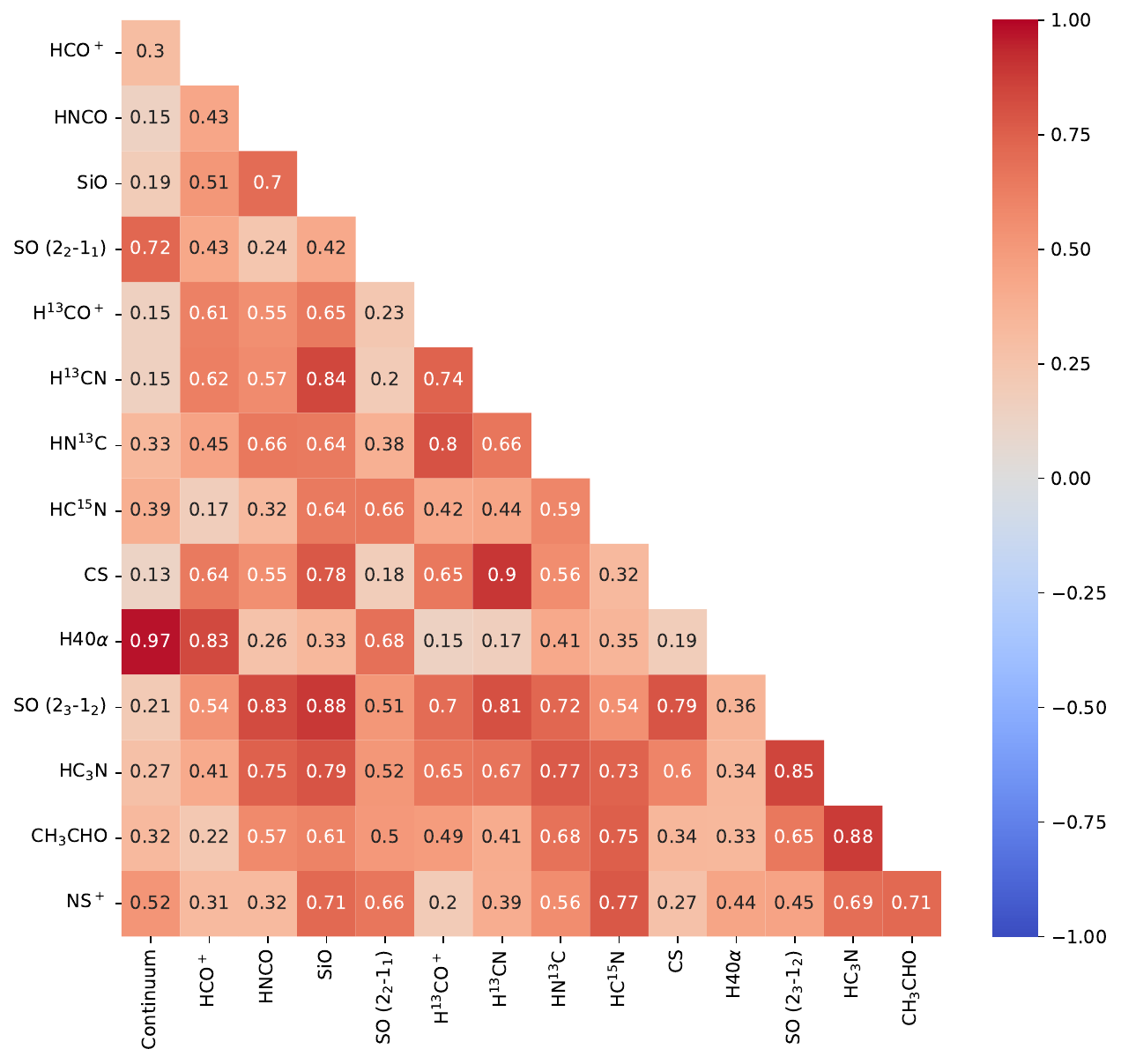}\\
\includegraphics[width=0.5\textwidth]{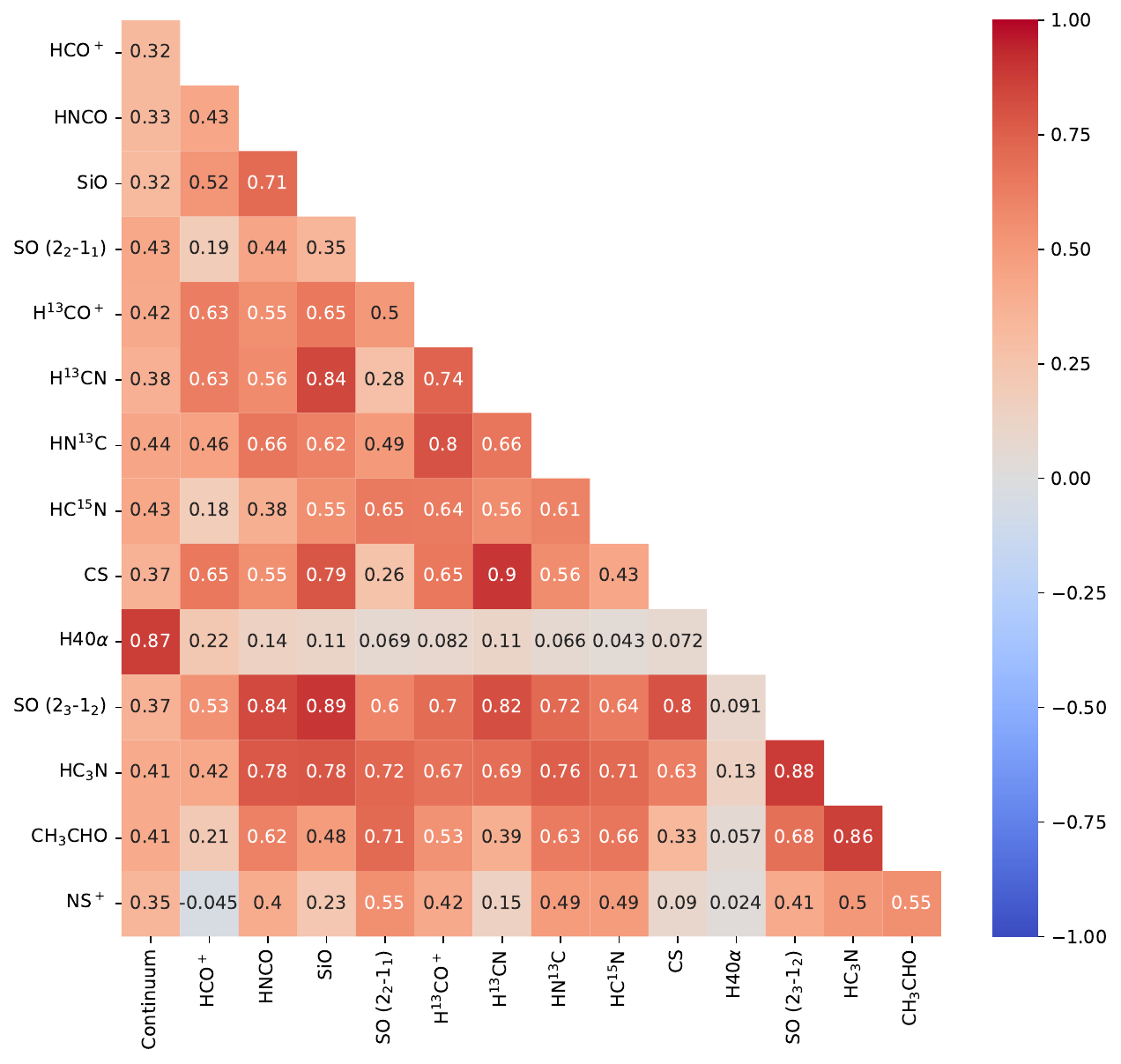}\\
\caption{Top: 2-D correlation matrix between the continuum and integrated intensities of the six molecular line emission presented in this paper as well as eight lines in the other SPWs \citepalias{Walker2026,Hsieh2026}. All the pixels in the ACES mosaic fields are considered. Bottom: the same correlation matrix but with pixels in Sgr~B2 and Sgr~A masked out. The rectangular mask towards Sgr~B2 is centered at (R.A., Dec.)=(17:47:20.14, $-$28\arcdeg{}22$'$54\farcs{66}), with a height of 26\farcs{15} and a height of 110\farcs{08}. The square mask towards Sgr~A is centered at (R.A., Dec.)=(17:45:40.14, $-$29\arcdeg{}00$'$28\farcs{24}), with a side length of 48\farcs{64}.
}
\label{fig:2dcorr}
\end{figure}

The 3~mm continuum emission of the survey has been presented in \citetalias{Ginsburg2026}, in which we have combined data from the ALMA 12-m array, the 7-m array, and the Green Bank Telescope MUSTANG-2 bolometer camera. Within the massive molecular clouds in the CMZ, e.g., Sgr~B2, Sgr~C, the 20~\kms{} cloud, and the Dust Ridge clouds, the continuum mostly originates from thermal dust emission in dense molecular gas, and therefore it is a good tracer of dense cores. Outside of the massive molecular clouds, the continuum could also stem from free-free emission in \hii{} regions or synchrotron emission.

The 2-dimensional (2-D) Pearson correlation coefficient ($\rho$) is a useful reference to reveal morphological similarities between different tracers and help understand their potential physical and chemical origins \citep[e.g.,][]{Lu2017,Callanan2023}. In \autoref{fig:2dcorr}, we present 2-D correlations between the continuum image and the integrated brightness images of all the major molecular lines covered in the ACES SPWs, including the six lines in 12-m SPWs 25 and 27. 
We calculate $\rho$ using the corrcoef function in NumPy. The resulting values of $\rho$ are between $-$1 and 1, with $-$1 being anti-correlated, 0 being not correlated at all, and 1 being correlated. Note that we present results with and without considering the Sgr~B2 and Sgr~A regions, where the continuum and line emissions are brightest among the ACES fields and therefore could bias the correlations.

Among the 14 lines, the radio recombination line H40$\alpha$ shows the highest correlation coefficient with the continuum. The commonly used dense gas tracers for the Galactic disk and external galaxies, including HNCO (4--3), \hcop{} (1--0), and CS (2--1), turn out to show only weak correlations with the continuum, probably because of a combined effect of the continuum not well tracing dense gas (e.g., free-free or synchrotron contamination) and the lines not correlating with dense gas (e.g., potentially optically thick line emission, enhancement of line emission by shocks). In the following, we only focus on the six lines covered by the 12-m SPWs 25 and 27, and leave in-depth discussions to future works.

When Sgr~B2 and Sgr~A are included in the calculation, the SO (2$_2$--1$_1$) line covered by the 12-m SPW 25 has the second highest correlation coefficient with the continuum ($\rho=0.72$) after H40$\alpha$, as shown in \autoref{fig:2dcorr} top panel. However, once Sgr~B2 is excluded, SO (2$_2$--1$_1$) only shows a weak correlation with the continuum ($\rho=0.43$; see \autoref{fig:2dcorr} bottom). We also find a strong correlation between SO and the continuum within the Sgr~B2 region ($\rho=0.71$). Excluding Sgr~A makes no difference to the correlation between the two.

Therefore, it is likely that the correlation between SO (2$_2$--1$_1$) and the continuum is biased by the bright emission in Sgr~B2. The SO (2$_2$--1$_1$) in Sgr~B2 has been mapped with single-dish observations by \citet{jones2008}, who noted that this line presents compact peaks near the two hot cores, Sgr~B2(M) and (N). This morphological similarity may be attributed to the fact that the SO (2$_2$--1$_1$) transition has the lowest Einstein coefficient and the highest upper level energy among the six lines (\autoref{tab:lines}), and therefore can be more easily excited in environments such as the hot cores in Sgr~B2. SO has been suggested to be enhanced by slow shocks ($\lesssim$20~\kms{}) that prevail in the CMZ \citep{Lis2001,Lu2017,Callanan2023}. It has been found to trace both dense molecular gas and shocks in other galactic nuclei \citep{Bouvier2024}; however, it is not clear if the SO emission actually traces unresolved protostellar outflows or shock-related filaments inside the dense gas \citep[e.g.,][]{Lu2021,Busch2024,Yang2025}. A combined analysis with other SO transitions as well as its isotopologue lines (e.g., SO (2$_3$--1$_2$), $^{33}$SO (2$_3$--1$_2$); see \citetalias{Hsieh2026}) covered in the other SPWs of the ACES observations will be helpful for understanding the excitation condition of this molecule.

The \hcfn{} (1--0) line has a weak correlation with the continuum, regardless of whether or not Sgr~B2 and Sgr~A are included in the analysis ($\rho=0.39$ or 0.44). Its low peak brightness temperature in \autoref{fig:mom8}, which is much lower than the typical kinetic temperature of 50--100~K \citep{Ginsburg2016,Krieger2017}, suggests that the line is optically thin in most of the surveyed area. Therefore, the \hcfn{} line could be a good tracer of dense gas in massive molecular clouds in this environment, even in Sgr~B2. Studies comparing the line emission and dense gas mass derived from \textit{Herschel} multi-wavelength observations can be carried out to confirm the robustness of this trend \citep[e.g.,][]{Mills2017b}.

The remaining four lines, including SiO (2--1), \htcop{} (1--0), \htcn{} (1--0), and \hntc{} (1--0), all have spatially diffuse and often filamentary emission across the whole CMZ. As such, their correlation coefficients with the spatially compact continuum are $\lesssim$0.3 when Sgr~B2 and Sgr~A are included, suggesting weak to no correlation. When Sgr~B2 and Sgr~A are excluded, their $\rho$ remains at $\sim$0.3--0.4, suggesting weak correlations. Between the four lines, however, we find strong correlations with $\rho>0.6$, which do not change when Sgr~B2 and Sgr~A are excluded in the correlation analysis. This result reflects their similarity in morphology. Indeed, the four lines have been found to be widespread in the CMZ in previous single dish observations \citep[e.g.,][]{Martin-Pintado1997,Riquelme2010,Jones2012,Tanaka2018}. The SiO emission is likely related to shocks in this environment \citep{Martin-Pintado1997,Liu2013,Lu2017,Coutens2017}. The \htcop{}, \htcn{}, and \hntc{} lines are usually used to trace molecular clouds given their high critical densities and their lower opacity with respect to their main isotopologues, HCO$^+$, HCN, and HNC \citep{Jones2012}. With the higher angular resolution of ACES, however, one could find that the three lines present ubiquitous filamentary structures across the CMZ, one of which can be clearly seen in the Sgr~B1 region, a feature similar to that seen in SiO. The filamentary structures could be related to active shock activities in the CMZ, whose origin is still under debate (e.g., expansion of \hii{} regions or supernova remnants: \citealt{Inutsuka2021,Henshaw2022}; cloud-cloud collision, tidal force, or shear: \citealt{Kruijssen2014a}; C.\ Battersby et al.\ 2025 submitted). This may suggest that the formation of the three isotopologues is related to shocks, e.g., they might be synthesized in the gas phase in post-shock regions. In such regions, shocks sputter SiO from dust grains and compress the gas to densities above 10$^5$~\cc{} \citep{Schilke1997,Bachiller1997}, which are consistent with the critical densities of the three isotopologue lines.

\subsection{Line ratios between \texorpdfstring{\htcn{}}{H13CN}, \texorpdfstring{\hntc{}}{HN13C}, and \texorpdfstring{\hcfn{}}{HC15N}}\label{subsec:dis_iso}

The relative intensities or densities of \htcn{} and \hcfn{} with respect to that of HCN are related to the $^{12}$C/$^{13}$C and $^{14}$N/$^{15}$N isotope abundance ratios, respectively. Lines of the main isotopologue H$^{12}$C$^{14}$N are not covered in the ACES observations. Nonetheless, the ratio between the \htcn{} and \hcfn{} lines is still informative. Previous observations toward nearby clouds have found that the integrated brightness temperature ratio between \htcn{} and \hcfn{} can vary between $\sim$3 and $\sim$10 \citep{Ikeda2002}. This translates to a ratio of their densities in the same range if one assumes the same excitation temperature and optically thin emission for both lines.

We present the ratio of integrated brightness temperatures of \hcfn{} (1--0) and \htcn{} (1--0) in the ACES data in \autoref{fig:lineratio}. Note that we do not derive the ratio of peak brightness temperatures of the two lines, because their spectral profiles are not the same and therefore their peaks do not align in the velocity axis. Comparing their integrated brightness temperatures has its own problem, e.g., the line widths of the two lines could be different. Nonetheless, we use the ratio in \autoref{fig:lineratio} as a starting point to demonstrate what could be learned from these two lines. Following the same assumptions in \citet{Ikeda2002}, the ratio is related to the compound isotope abundance ratio ($^{12}$C/$^{13}$C)($^{14}$N/$^{15}$N).

If the $^{12}$C/$^{13}$C isotope ratio could be estimated, e.g., from other isotopologue pairs \citepalias[\htcop{} and HCO$^+$; for the latter line, see][]{Walker2026}, then we can decouple the nitrogen isotopic ratio $^{14}$N/$^{15}$N from the compound ratio derived from \hcfn{} and \htcn{}. The nitrogen isotopic ratio has been shown to vary significantly in different environments \citep{Fontani2015,Colzi2018}, and is an important constraint in astrochemical models and in simulations of the evolution of the Galactic stellar population.

The \hntc{} molecule is an isotopomer of \htcn{}, and their line ratio is suggested to be anti-correlated with the gas temperature and therefore a potential gas thermometer \citep[e.g.,][]{Hirota1998,Jin2015,Pazukhin2022}. Both lines are optically thinner than their respective main isotopologue HNC and HCN, whose ratio has been shown to be correlated with the gas temperature \citep{Graninger2014,Hacar2020} but may be biased when the emission is optically thick. Therefore, the \hntc{}/\htcn{} line ratio is particularly of interest in the CMZ, where the HCN emission becomes optically thick \citep{Jones2012}.

As shown in \autoref{fig:lineratio}, the \hntc{}/\htcn{} line ratio varies between $\sim$0.01 and $>$10 (truncated in the plot). The line ratio shows a possible increasing trend towards the interiors of molecular clouds including Sgr~B2, the Brick, and the 20~\kms{} cloud. Clouds in the CMZ are known to be externally heated \citep[e.g.,][]{Molinari2011,Rathborne2014b}. Therefore, the anti-correlation between the line ratio and temperature could still hold for the CMZ. With the high resolution of the ACES observations, one will be able to explore the \hntc{}/\htcn{} line ratio and the gas temperature derived from e.g., multiple NH$_3$ lines across various environments in the CMZ. This will help sort out the impact of other factors on the ratio such as the ultraviolet radiation \citep{Santa-Maria2023,Harada2024} and calibrate this ratio as a gas thermometer, which then can be applied to studies of external galaxies \citep[e.g.,][]{Behrens2022,Harada2024}.

Finally, in \autoref{fig:lineratio} we also present the line ratio map between \htcop{} (1--0) and \hcop{} (1--0), the latter of which is released in \citetalias{Walker2026}. This ratio can be used as a metric of the $^{12}$C/$^{13}$C isotope ratio, although \hcop{} likely becomes optically thick toward dense molecular clouds. As such, peaks in the bottom panel of \label{fig:lineratio}, e.g., towards Sgr~B2, may not necessarily suggest high $^{12}$C/$^{13}$C isotope ratios, but simply optically thick \hcop{} emission. In-depth analysis of this ratio is beyond the scope of this paper.

\begin{figure*}
\centering
\includegraphics[width=1.0\textwidth]{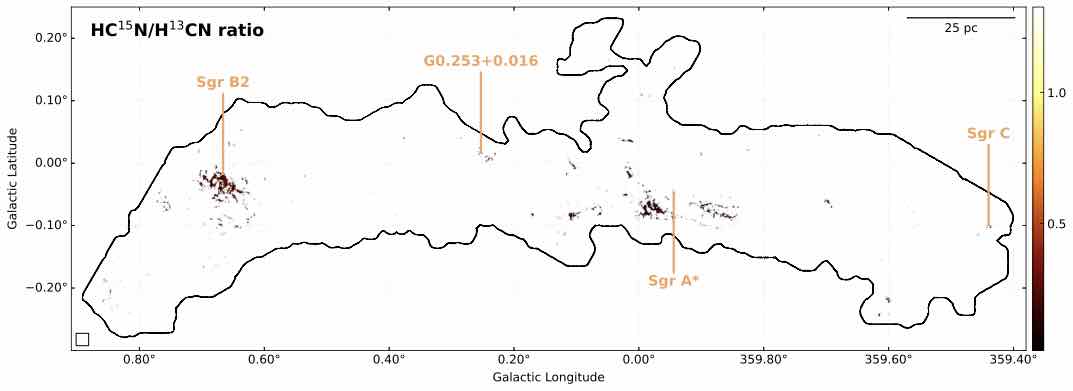}
\includegraphics[width=1.0\textwidth]{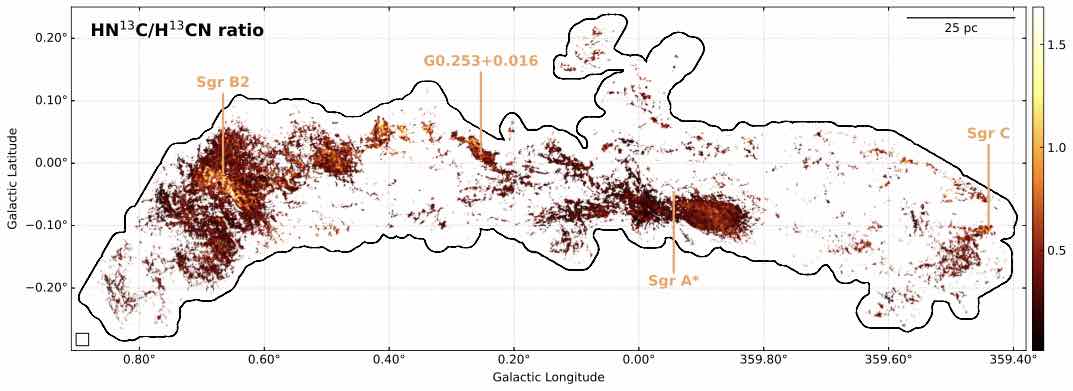}
\includegraphics[width=1.0\textwidth]{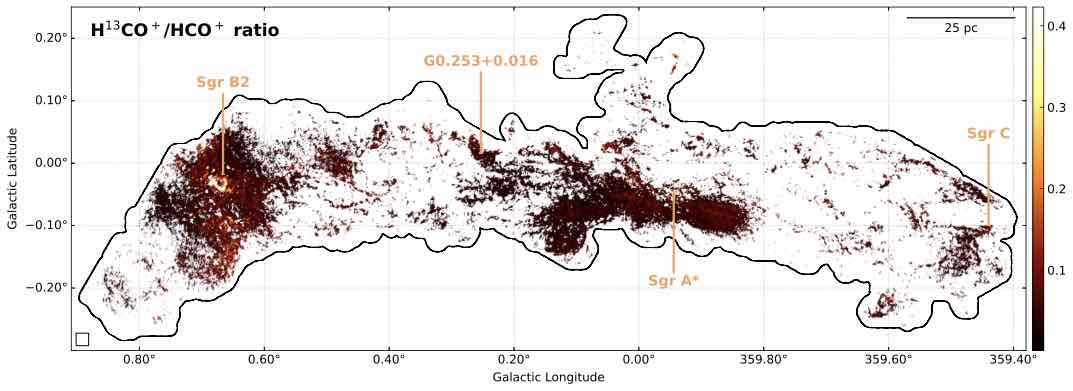}
\caption{Ratios of integrated brightness temperatures between \hcfn{} 1--0 and \htcn{} 1--0 (top), \hntc{} 1--0 and \htcn{} 1--0 (middle), and \htcop{} 1--0 and \hcop{} 1--0 (bottom), respectively. The color bars are truncated at the 99.9 percentile of the maximum value.}
\label{fig:lineratio}
\end{figure*}

\subsection{Other interesting features in the data}\label{subsec:dis_other}

The maps in Figures~\ref{fig:mom0}--\ref{fig:mom1-continued} reveal prominent morphological features in the CMZ, including rings, filaments, and clumpy clouds. Here we highlight a few features of interest, and leave detailed discussions to science papers that are currently being prepared by the ACES team.

\citet{Nonhebel2024} have reported a detailed study of the M0.8$-$0.2 ring using the ACES data, which can be seen in the maps of SiO, \htcn{}, \htcop{}, and \hntc{}. The molecular line data are used for constraining the kinematics and energetics of the ring. There are numerous other ring-like structures of different angular scales in the data, which might be related to supernova remnants or \hii{} regions \citep[e.g., `atolls' in][]{Hatchfield2024}. A catalog of such objects will be valuable for studying the star formation history of the CMZ.

Filamentary structures at spatial scales of 0.1~pc to $\gtrsim$10~pc are known to be ubiquitous in molecular line emission in the CMZ \citep[e.g.,][]{Bally2014,Johnston2014,Henshaw_2019,Wallace2022,Yang2025}. A large population of filaments are revealed in the molecular line emission presented in \autoref{sec:data}, particularly towards the Sgr~B1 region between Galactic longitudes of 0\arcdeg{}20$'$--0\arcdeg{}40$'$ and latitudes of 0\arcdeg{}--$-$0\arcdeg{}5$'$, which may trace the photo-dissociation region between the young massive star cluster Sgr~B1 and the surrounding molecular gas \citep{Simpson2018a,Simpson2021}. The exact nature of the filaments, which could be diverse in spite of their similar morphologies, is yet to be explored with the ACES data (e.g., C.\ Battersby et al. 2026 submitted).

\section{Conclusions}
\label{sec:conclusions}

In this paper, we have presented an overview of spectral line data of the two intermediate-width spectral windows from the ACES Large Program, with a focus on the six bright lines including SiO (2--1), SO (2$_2$--1$_1$), \htcop{} (1--0), \htcn{} (1--0), \hntc{} (1--0), and \hcfn{} (1--0). We present the integrated brightness temperature, the peak brightness temperature, the centroid velocity, and the position-velocity diagrams along the Galactic longitude of the six lines. The data products are released along with the paper.

In addition, we showcase potential science that could be done with the data, including morphological correlations between the 3~mm continuum \citepalias{Ginsburg2026} and the spectral line emissions (\citetalias{Walker2026,Hsieh2026}, and this work) to understand the excitation conditions and chemical properties of the molecules, line ratios between isotopologues and isomers to estimate the isotope abundance ratio, and other features in the data such as rings and filaments. We leave in-depth discussions of these topics to follow-up papers in the ACES series and to users of the data.

\section*{Acknowledgements}
The paper was instigated and led by the ACES data reduction working group, which is coordinated by Adam Ginsburg, Daniel Walker, and Ashley Barnes, and includes (alphabetically) Nazar Budaiev, Laura Colzi, Claire Cooke, Savannah Gramze, Pei-Ying Hsieh, Desmond Jeff, Xing Lu, Jaime Pineda, Marc Pound, \'{A}lvaro S\'{a}nchez-Monge, and Qizhou Zhang.
ACES is led by Principal Investigator Steven Longmore, together wwith co-PIs and the management team, John Bally, Ashley Barnes, Cara Battersby, Laura Colzi, Adam Ginsburg, Jonathan Henshaw, Paul Ho, Izaskun Jim\'{e}nez-Serra, Elisabeth Mills, Maya Petkova, Mattia Sormani, Robin Tress, Daniel Walker, and Jennifer Wallace. 
The ACES proposal was led by Principal Investigator Steven Longmore, together with co-PIs John Bally, Cara Battersby,  Adam Ginsburg, Jonathan Henshaw, Paul Ho, J.\ M.\ Diederik Kruijssen,  Izaskun Jim\'{e}nez-Serra, and Elisabeth Mills. 
The remaining coauthors comprise the ACES team, and its members contributed to the proposal, data analysis, idea development, and/or reading and commenting on the manuscript.
X.L.\ acknowledges support from the Strategic Priority Research Program of the Chinese Academy of Sciences (CAS) Grant No.\ XDB0800300, the National Key R\&D Program of China (No.\ 2022YFA1603101), the National Natural Science Foundation of China (NSFC) through grant Nos.\ 12273090 and 12322305, the Natural Science Foundation of Shanghai (No.\ 23ZR1482100), and the CAS ``Light of West China'' Program No.\ xbzg-zdsys-202212.
D.L.W gratefully acknowledges support from the UK ALMA Regional Centre (ARC) Node, which is supported by the Science and Technology Facilities Council [grant numbers ST/Y004108/1 and ST/T001488/1].
R.F. acknowledges support from the grants PID2023-146295NB-I00, and from the Severo Ochoa grant CEX2021-001131-S funded by MCIN/AEI/ 10.13039/501100011033 and by ``European Union NextGenerationEU/PRTR''. 
COOL Research DAO \citep{cool_whitepaper} is a Decentralized Autonomous Organization supporting research in astrophysics aimed at uncovering our cosmic origins.
I.J-.S., L.C., and V.M.R. acknowledge support from the grant PID2022-136814NB-I00 by the Spanish Ministry of Science, Innovation and Universities/State Agency of Research MICIU/AEI/10.13039/501100011033 and by ERDF, UE. I.J-.S. also acknowledges the ERC Consolidator grant OPENS (project number 101125858) funded by the European Union. V.M.R. also acknowledges the grant RYC2020-029387-I funded by MICIU/AEI/10.13039/501100011033 and by "ESF, Investing in your future", and from the Consejo Superior de Investigaciones Cient{\'i}ficas (CSIC) and the Centro de Astrobiolog{\'i}a (CAB) through the project 20225AT015 (Proyectos intramurales especiales del CSIC); and from the grant CNS2023-144464 funded by MICIU/AEI/10.13039/501100011033 and by “European Union NextGenerationEU/PRTR.
The authors acknowledge UFIT Research Computing for providing computational resources and support that have contributed to the research results reported in this publication. 
A.G acknowledges support from the NSF under grants CAREER 2142300, AAG 2008101, and particularly AAG 2206511 that supports the ACES large program.
P.-Y. H. acknowledges support from the ADC. Data analysis was in part carried out on the Multi-wavelength Data Analysis System operated by the Astronomy Data Center (ADC), National Astronomical Observatory of Japan.
E.A.C.\ Mills  gratefully  acknowledges  funding  from the National  Science  Foundation  under  Award  Nos. 1813765, 2115428, 2206509, and CAREER 2339670.
M.G.S.-M.\ acknowledges support from the NSF under grant CAREER 2142300. M.G.S.-M.\ also thanks the Spanish MICINN for funding support under grant PID2023-146667NB-I00.
A.S.-M.\ acknowledges support from the RyC2021-032892-I grant funded by MCIN/AEI/10.13039/501100011033 and by the European Union `Next GenerationEU’/PRTR, as well as the program Unidad de Excelencia María de Maeztu CEX2020-001058-M, and support from the PID2023-146675NB-I00 (MCI-AEI-FEDER, UE).
K.M.D acknowledges support from the European Research Council (ERC) Advanced Grant MOPPEX 833460.vii.
Q.Z. gratefully acknowledges the support from the National Science Foundation under Award No. AST-2206512, and the Smithsonian Institute FY2024 Scholarly Studies Program.
FHL acknowledges support from the ESO Studentship Programme, the Scatcherd European Scholarship of the University of Oxford, and the European Research Council’s starting grant ERC StG-101077573 (`ISM-METALS').
C.F.~acknowledges funding provided by the Australian Research Council (Discovery Projects DP230102280 and DP250101526), and the Australia-Germany Joint Research Cooperation Scheme (UA-DAAD).
F. M acknowledges financial support from the School of Astronomy at the Institute for Research in Fundamental Sciences-IPM.
J.K. is supported by the Royal Society under grant number RF\textbackslash ERE\textbackslash231132, as part of project URF\textbackslash R1\textbackslash211322.
M.C.\ gratefully acknowledges funding from the DFG through an Emmy Noether Research Group (grant number CH2137/1-1).
MCS acknowledges financial support from the European Research Council under the ERC Starting Grant ``GalFlow'' (grant 101116226) and from Fondazione Cariplo under the grant ERC attrattivit\`{a} n. 2023-3014.
J.Wallace gratefully acknowledges funding from National Science Foundation under Award Nos. 2108938 and 2206510. 
F.N.-L. gratefully acknowledges financial support from grant PID2024-162148NA-I00, funded by MCIN/AEI/10.13039/501100011033 and the European Regional Development Fund (ERDF) “A way of making Europe”, from the Ramón y Cajal programme (RYC2023-044924-I) funded by MCIN/AEI/10.13039/501100011033 and FSE+, and from the Severo Ochoa grant CEX2021-001131-S, funded by MCIN/AEI/10.13039/501100011033.
C.B.\ gratefully acknowledges funding from National Science Foundation under Award Nos.\ 2108938, 2206510, and CAREER 2145689, as well as from the National Aeronautics and Space Administration through the Astrophysics Data Analysis Program under Award ``3-D MC: Mapping Circumnuclear Molecular Clouds from X-ray to Radio,” Grant No.\ 80NSSC22K1125.
S.Z.\ acknowledges support from the NAOJ ALMA Scientific Research Grant Code 2025-29B.
This paper makes use of the following ALMA data: ADS/JAO.ALMA\#2021.1.00172.L. ALMA is a partnership of ESO (representing its member states), NSF (USA) and NINS (Japan), together with NRC (Canada), NSTC and ASIAA (Taiwan), and KASI (Republic of Korea), in cooperation with the Republic of Chile. The Joint ALMA Observatory is operated by ESO, AUI/NRAO and NAOJ.
The authors are grateful to the staff throughout the ALMA organisation, particularly those at the European ALMA Regional Centre, the Joint ALMA Observatory, and the UK ALMA Regional Centre Node, for their extensive support, which was essential to the success of this challenging Large Program.
The authors thank the anonymous referee, whose feedback was critical for improving the clarity and utility of this paper.

\section*{Software}
The ACES pipeline is based on a number of open-source astronomy software packages, including \texttt{astropy} \citep{astropy:2013, astropy:2018, astropy:2022}, \texttt{astroquery} \citep{Ginsburg2019astroquery}, \texttt{spectral-cube} \citep{Ginsburg2019spectralcube}, \texttt{radio-beam} \citep{Koch2025radiobeam}, pvextractor \citep{Ginsburg2016pvextractor}, \texttt{reproject} \citep{Robitaille2020reproject}, \texttt{statcont} \citep{Sanchez-Monge2018}, CASA \citep{CASA2022}, numpy \citep{Harris2020}, and matplotlib \citep{Hunter2007}. In addition, this paper made use of the following software packages: CARTA \citep{carta} to visualise and analyse the data, and APLpy \citep{aplpy} and seaborn \citep{Waskom2021} to make plots.

\section*{Data Availability}
All data products and associated documentation can be found at \url{https://almascience.org/alma-data/lp/aces}. To maximise usability and provide a single, comprehensive resource for users, we also provide an online, machine-readable megatable, which contains the full, detailed list of all data products released for the entire survey, including hyperlinks to the full files from the ALMA archive.

All code and data processing issues are available at the public GitHub repository here: \url{https://github.com/ACES-CMZ/reduction_ACES}.

The ACES data reduction was a gigantic monolithic work that was written up in 5 papers.  If you use the ACES data, please cite the appropriate works, which includes \citet{Longmore2026} and the data papers: continuum \citep{Ginsburg2026} and cubes in high-resolution \citep{Walker2026}, medium-resolution (this work), and low-resolution \citep{Hsieh2026} spectral windows. Note that all papers should be cited for use of any data; the continuum imaging relied on the line papers, and vice-versa.

\bibliographystyle{mnras}
\bibliography{refs}

\section*{Author Affiliations}
\printaffiliation{shao}{Shanghai Astronomical Observatory, Chinese Academy of Sciences, 80 Nandan Road, Shanghai 200030, P.\ R.\ China}
\printaffiliation{naoc_key}{State Key Laboratory of Radio Astronomy and Technology, A20 Datun Road, Chaoyang District, Beijing, 100101, P. R. China}
\printaffiliation{ukarcnode}{UK ALMA Regional Centre Node, Jodrell Bank Centre for Astrophysics, The University of Manchester, Manchester M13 9PL, UK}
\printaffiliation{uflorida}{Department of Astronomy, University of Florida, P.O. Box 112055, Gainesville, FL 32611, USA}
\printaffiliation{eso}{European Southern Observatory (ESO), Karl-Schwarzschild-Stra{\ss}e 2, 85748 Garching, Germany}
\printaffiliation{naoj}{National Astronomical Observatory of Japan, 2-21-1 Osawa, Mitaka, Tokyo 181-8588, Japan}
\printaffiliation{ice_csic}{Institut de Ci\`encies de l'Espai (ICE), CSIC, Campus UAB, Carrer de Can Magrans s/n, E-08193, Bellaterra, Barcelona, Spain}
\printaffiliation{ieec}{Institut d'Estudis Espacials de Catalunya (IEEC), E-08860, Castelldefels, Barcelona, Spain}
\printaffiliation{umd}{University of Maryland, Department of Astronomy, College Park, MD 20742-2421, USA}
\printaffiliation{mpe}{Max-Planck-Institut f\"ur extraterrestrische Physik, Gie\ss enbachstra\ss e 1, 85748 Garching bei M\"unchen, Germany}
\printaffiliation{kansas}{Department of Physics and Astronomy, University of Kansas, 1251 Wescoe Hall Drive, Lawrence, KS 66045, USA}
\printaffiliation{ljmu}{Astrophysics Research Institute, Liverpool John Moores University, 146 Brownlow Hill, Liverpool L3 5RF, The UK}
\printaffiliation{mpia}{{Max Planck Institute for Astronomy, K\"{o}nigstuhl 17, D-69117 Heidelberg, Germany}}
\printaffiliation{ari_heidelberg}{Astronomisches Rechen-Institut, Zentrum f\"{u}r Astronomie der Universit\"{a}t Heidelberg, M\"{o}nchhofstra\ss e 12-14, D-69120 Heidelberg, Germany}
\printaffiliation{COOL}{Cosmic Origins Of Life (COOL) Research DAO, \href{https://coolresearch.io}{https://coolresearch.io}}
\printaffiliation{eso_chile}{European Southern Observatory, Alonso de C\'ordova, 3107, Vitacura, Santiago 763-0355, Chile}
\printaffiliation{jao}{Joint ALMA Observatory, Alonso de C\'ordova, 3107, Vitacura, Santiago 763-0355, Chile}
\printaffiliation{nanjing}{School of Astronomy and Space Science, Nanjing University, 163 Xianlin Avenue, Nanjing 210023, P.R.China}
\printaffiliation{nanjing_key}{Key Laboratory of Modern Astronomy and Astrophysics (Nanjing University), Ministry of Education, Nanjing 210023, P.R.China}
\printaffiliation{cfa}{Center for Astrophysics | Harvard \& Smithsonian, 60 Garden Street, Cambridge, MA 02138, USA}
\printaffiliation{colorado}{Center for Astrophysics and Space Astronomy, Department of Astrophysical and Planetary Sciences, University of Colorado, Boulder, CO 80389, USA}
\printaffiliation{uconn}{Department of Physics, University of Connecticut, 196A Auditorium Road, Unit 3046, Storrs, CT 06269, USA}
\printaffiliation{cab_csic}{Centro de Astrobiolog{\'i}a (CAB), CSIC-INTA, Carretera de Ajalvir km 4, 28850 Torrej{\'o}n de Ardoz, Madrid, Spain}
\printaffiliation{iaa_taipei}{Academia Sinica Institute of Astronomy and Astrophysics, Astronomy-Mathematics Building, AS/NTU No.1, Sec. 4, Roosevelt Rd, Taipei 10617, Taiwan}
\printaffiliation{chalmers}{Space, Earth and Environment Department, Chalmers University of Technology, SE-412 96 Gothenburg, Sweden}
\printaffiliation{clap}{{Como Lake centre for AstroPhysics (CLAP), DiSAT, Universit{\`a} dell’Insubria, via Valleggio 11, 22100 Como, Italy}}
\printaffiliation{iop_epfl}{Institute of Physics, Laboratory for Galaxy Evolution and Spectral Modelling, EPFL, Observatoire de Sauverny, Chemin Pegasi 51, 1290 Versoix, Switzerland}
\printaffiliation{oaq}{Observatorio Astron\'omico de Quito, Observatorio Astron\'omico Nacional, Escuela Polit\'ecnica Nacional, 170403, Quito, Ecuador}
\printaffiliation{inaf_arcetri}{INAF Arcetri Astrophysical Observatory, Largo Enrico Fermi 5, Firenze, 50125, Italy}
\printaffiliation{jbca}{Jodrell Bank Centre for Astrophysics, The University of Manchester, Manchester M13 9PL, UK}
\printaffiliation{nrao}{National Radio Astronomy Observatory, 520 Edgemont Road, Charlottesville, VA 22903, USA}
\printaffiliation{ita_heidelberg}{Universit\"{a}t Heidelberg, Zentrum f\"{u}r Astronomie, Institut f\"{u}r Theoretische Astrophysik, Albert-Ueberle-Str 2, D-69120 Heidelberg, Germany}
\printaffiliation{anu}{Research School of Astronomy and Astrophysics, Australian National University, Canberra, ACT 2611, Australia}
\printaffiliation{iaa_csic}{Instituto de Astrof\'{i}sica de Andaluc\'{i}a, CSIC, Glorieta de la Astronomía s/n, 18008 Granada, Spain}
\printaffiliation{ucn}{Instituto de Astronom\'ia, Universidad Cat\'olica del Norte, Av. Angamos 0610, Antofagasta, Chile}
\printaffiliation{cassaca}{Chinese Academy of Sciences South America Center for Astronomy, National Astronomical Observatories, CAS, Beijing 100101, China}
\printaffiliation{ias}{Institute for Advanced Study, 1 Einstein Drive, Princeton, NJ 08540, USA}
\printaffiliation{ucl}{Department of Physics and Astronomy, University College London, Gower Street, London WC1E 6BT, UK}
\printaffiliation{izw_heidelberg}{Universit\"{a}t Heidelberg, Interdisziplin\"{a}res Zentrum f\"{u}r Wissenschaftliches Rechnen, Im Neuenheimer Feld 225, 69120 Heidelberg, Germany}
\printaffiliation{ipm}{Institute for Research in Fundamental Sciences (IPM), School of Astronomy, Tehran, Iran}
\printaffiliation{ulaserena}{Departamento de Astronom\'ia, Universidad de La Serena, Ra\'ul Bitr\'an 1305, La Serena, Chile}
\printaffiliation{iff_csic}{Instituto de Física Fundamental (CSIC), Calle Serrano 121-123, 28006, Madrid, Spain}
\printaffiliation{gbo}{Green Bank Observatory, P.O. Box 2, Green Bank, WV 24944, USA}
\printaffiliation{utokyo}{Institute of Astronomy, The University of Tokyo, Mitaka, Tokyo 181-0015, Japan}
\printaffiliation{umass}{Department of Astronomy, University of Massachusetts, Amherst, MA 01003, USA}
\printaffiliation{aberystwyth}{Department of Physics, Aberystwyth University, Ceredigion, Cymru, SY23 3BZ, UK}
\printaffiliation{kiaa_pku}{Kavli Institute for Astronomy and Astrophysics, Peking University, Beijing 100871, People's Republic of China}
\printaffiliation{pku_astro}{Department of Astronomy, School of Physics, Peking University, Beijing, 100871, People's Republic of China}
\printaffiliation{ist}{Department of Earth and Planetary Sciences, Institute of Science Tokyo, Meguro, Tokyo, 152-8551, Japan}

\onecolumn
\appendix

\section{Data release summary}\label{subsec:data_appendix}
\noindent
The data products from the ACES survey follow a uniform naming convention. The filenames for the data products presented in this paper can be constructed using the templates described below, in conjunction with the information provided in Appendix A of \citet{Walker2026}. 

The following subsections are divided according to different product types which occur at the Member ObsUnitSet (MOUS) and Group ObsUnitSet (GOUS) levels\footnote{For a description of ALMA project structures and naming conventions, please refer to Section 2 of the Cycle 9 QA2 guide: \url{https://almascience.eso.org/documents-and-tools/cycle9/alma-qa2-data-products-for-cycle-9}}.
\\
\subsection{Member-level products (12m and 7m)}
We release the data cubes for all SPWs for the 45 ACES regions. These are member-level products, and correspond to the cubes for the 12m and 7m arrays for each field prior to combination. 

We release the cubes for the separate arrays as they are different to those initially delivered by ALMA due to the changes that were made during reprocessing. We do not re-release any of the stand-alone TP products, as we used the same cubes as those available in the ALMA Science Archive (ASA).

In this and subsequent subsections, we describe how to construct the filenames of the released data products by providing filename templates that include placeholder variables. These placeholders can be changed as needed to build the filename of the relevant product.

Following the requirements of the ASA, the filename template is:

\texttt{member.uid\_\_\_A001\_\{\textbf{mous-id}\}.lp\_2021.1.00172.L.slongmore.\{\textbf{field\_coords}\}.\{\textbf{array}\}.\{\textbf{frqrange}\}.cube.pbcor.fits}

The placeholder template components should be replaced as follows:
\begin{itemize}
    \item \texttt{\{\textbf{mous-id}\}}: The 12m or 7m MOUS ID for a given field, as listed in Table A1 in \citet{Walker2026}.
    \item \texttt{\{\textbf{field\_coords}\}}: The central coordinates of the field, as listed in Table A1 in \citet{Walker2026} (e.g., \texttt{G000.073+0.184}).
    \item \texttt{\{\textbf{array}\}}: The ALMA array, either \texttt{12m} or \texttt{7m}.
    \item \texttt{\{\textbf{frqrange}\}}: The frequency range, which depends on the SPW:
    \begin{itemize}
        \item SPW 25: \texttt{86.0-86.4GHz}
        \item SPW 27: \texttt{86.7-87.1GHz}
    \end{itemize}
\end{itemize}

For example, to find the 12m-only cube for the 12-m SPW 25 for field \texttt{ao} (this field contains the well-known molecular cloud G0.253+0.016, aka the Brick), the resulting filename would be:

\texttt{member.uid\_\_\_A001\_X15a0\_X190.lp\_2021.1.00172.L.slongmore.G000.300+0.063.12m.86.0-86.4GHz.cube.pbcor.fits}

\subsection{Group-level products (Combined cubes per region)}

We also release array-combined products per SPW per region. As these products combine data from multiple MOUSs, they are deemed to be group-level products. The filenames of these products can be constructed as follows:

\texttt{group.uid\_\_\_A001\_X1590\_X30a9.lp\_2021.1.00172.L.slongmore.\{\textbf{field\_coords}\}.\{\textbf{arrays}\}.\{\textbf{frqrange}\}.cube.pbcor.fits}

Where the template components are:
\begin{itemize}
    \item \texttt{\{\textbf{field\_coords}\}} and \texttt{\{\textbf{frqrange}\}}: follow the same definition as for the member-level products.
    \item \texttt{\{\textbf{arrays}\}}: \texttt{12m7mTP}
\end{itemize}

For example, the corresponding 12-m SPW 25 cube including 12m, 7m, and TP data for field \texttt{ao} would be:

\texttt{group.uid\_\_\_A001\_X1590\_X30a9.lp\_2021.1.00172.L.slongmore.G000.300+0.063.12m7mTP.86.0-86.4GHz.cube.pbcor.fits}

A summary of the combined cubes for the two intermediate-width SPWs, including the original beam shapes, pixel sizes, and imaging rms of the individual fields, is reported in Tables~\ref{tab:cubestats25} and \ref{tab:cubestats27}.

\subsection{Group-level products (Full CMZ mosaics)}

Finally, we release the full, contiguous mosaics covering the entire ACES footprint. Again, since these result from the combination of many MOUSs, the full mosaics are at the group level.

In contrast to cubes for the individual fields, we do not release full mosaics for all SPWs, but rather for the 6 specific lines given in \autoref{tab:lines}.
The SPWs are sufficiently broad that they contain significant spectral ranges with no line emission. We therefore opted to not produce full mosaics containing these blank channels in the interest of reducing the file sizes. In addition to the data cubes for each line, we also release a suite of advanced products, including moment maps, noise maps, etc. (see below).

The filename template for the full mosaic cubes and associated products is:

\texttt{group.uid\_\_\_A001\_X1590\_X30a9.lp\_2021.1.00172.L.slongmore.cmz\_mosaic.\{\textbf{arrays}\}.\{\textbf{molecule}\}.\{\textbf{suffix}}\}

Where the template components are:
\begin{itemize}
    \item \texttt{\{\textbf{arrays}\}}: \texttt{12m7mTP} 
    \item \texttt{\{\textbf{molecule}\}}: \texttt{H13CN, H13COplus, HC15N, HN13C, SiO21, or SO21}.
    \item \texttt{\{\textbf{suffix}\}} defines the specific data product. Products released with this paper are:
    \begin{itemize}
        \item \texttt{cube.pbcor.fits}: primary beam corrected image cube.
        \item \texttt{cube.downsampled\_spatially.pbcor.fits}: as previous, spatially smoothed to a 5 arcsec beam and then spatially rebinned by a factor of $9\times9$.
        \item \texttt{integrated\_intensity.fits}: masked integrated intensity map of the full-resolution cube.
        \item \texttt{mad\_std.fits}: noise map of the full-resolution cube.
        \item \texttt{peak\_intensity.fits}: peak intensity map of the full-resolution cube.
        \item \texttt{velocity\_at\_peak\_intensity.fits}: velocity map from the masked full-resolution cube; velocity estimated using the peak intensity of the spectrum for each pixel.
        \item \texttt{PV\_l\_max.fits}: $\ell$–$v$ position–velocity map from the full-resolution cube; maximum intensity taken along Galactic latitude ($b$).
        \item \texttt{PV\_l\_mean.fits}: $\ell$–$v$ position–velocity map from the masked full-resolution cube; mean intensity taken along Galactic latitude ($b$)
        \item \texttt{PV\_b\_max.fits}: $b$–$v$ position–velocity map from the full-resolution cube; maximum intensity taken along Galactic longitude ($\ell$).
        \item \texttt{PV\_b\_mean.fits}: $b$–$v$ position–velocity map from the full-resolution cube; mean intensity taken along Galactic longitude ($\ell$).                
    \end{itemize}
\end{itemize}

As an example, the peak $\ell$–$v$ map for the full ACES mosaic of the combined SiO (2--1) data would be constructed as:

\texttt{group.uid\_\_\_A001\_X1590\_X30a9.lp\_2021.1.00172.L.slongmore.cmz\_mosaic.12m7mTP.SiO21.PV\_l\_max.fits}

\begin{table*}
    \centering
    \caption{Global statistics for the 12-m SPW 25 cubes, after feathering data from the 12-m, 7-m, and TP arrays altogether, for all ACES fields. The noise is estimated from channels in the lowest quartile of the mean spectrum (assumed emission-free) using the scaled median absolute deviation (${\rm MAD}$; where $\sigma_{\rm MAD} = 1.4826\,{\rm MAD}$).}
    \label{tab:cubestats25}
    \begin{tabular}{cccccccccccc}
        \hline \hline
        Field & V$_{\textrm{off}}$ & B$_{\textrm{maj}}$ & B$_{\textrm{min}}$ & BPA & Image size & No.\ channels & Pixel size & S$_\text{peak}$ & $\sigma_{\rm MAD}$ & S$_\text{peak}$ & $\sigma_{\rm MAD}$ \\
        & (\kms) & (\arcsec) & (\arcsec) & (\arcdeg) & (pix) & & (\arcsec) & (\mjypbm{}) & (\mjypbm{}) & (K) & (K) \\
        \hline
		a & 40  & 1.69 & 1.20 & $-$77 & (2094, 2499) & 1912 & 0.20 & 454.3	& 5.2 & 36.9  & 0.4 \\
        aa & 0  & 2.04 & 1.74 & $-$75 & (1623, 1481) & 1917 & 0.26 & 1527.0& 4.7 & 70.9  & 0.2 \\
        ab & 0  & 1.99 & 1.70 & 88	 & (1496, 1261) & 1915 & 0.28 & 729.3	& 4.5 & 35.5  & 0.2 \\
        ac & 0  & 1.78 & 1.20 & $-$78 & (1686, 1626) & 1914 & 0.19 & 987.6	& 5.2 & 76.2  & 0.4 \\
        ad & 0  & 2.02 & 1.19 & $-$75 & (1184, 2564) & 1920 & 0.20 & 668.7	& 5.7 & 45.8  & 0.4 \\
        ae & 0  & 2.26 & 1.84 & 83	 & (1343, 1365) & 1914 & 0.30 & 849.2	& 4.9 & 33.6  & 0.2 \\
        af & 0  & 2.79 & 1.71 & 89	 & (1446, 1176) & 1915 & 0.31 & 1009.9	& 5.0 & 34.9  & 0.2 \\
        ag & 0  & 1.91 & 1.46 & $-$79 & (1731, 1721) & 1912 & 0.24 & 6629.9& 5.0 & 391.6 & 0.3 \\
        ah & 30 & 1.87 & 1.20 & $-$72 & (2004, 1942) & 1912 & 0.20 & 600.8	& 4.4 & 44.1  & 0.3 \\
        ai & 0  & 2.55 & 1.77 & 78	 & (1378, 1622) & 1914 & 0.29 & 1217.4	& 4.6 & 44.4  & 0.2 \\
        aj &$-$30 & 2.45 & 1.79 & 85	 & (1439, 1288) & 1914 &0.31&20814.4& 4.4 & 781.8 & 0.2 \\
        ak & 30 & 2.48 & 2.11 & 82	 & (764, 1009)  & 1916 & 0.32 & 448.6	& 4.8 & 14.1  & 0.2 \\
        al & 0  & 2.77 & 2.04 & 86	 & (2190, 1447) & 1920 & 0.28 & 679.6	& 5.8 & 19.8  & 0.2 \\
        am & 30 & 3.14 & 1.98 & $-$85 & (2612, 1140) & 1920 & 0.30 & 622.9	& 6.3 & 16.5  & 0.2 \\
        an & 30 & 2.55 & 1.73 & 83	 & (1335, 1359) & 1914 & 0.30 & 4246.9	& 5.1 & 158.6 & 0.2 \\
        ao & 0  & 2.52 & 1.62 & $-$86 & (2512, 923)	& 1914 & 0.28&2742.1& 6.0 & 110.6 & 0.2 \\
        ap & 0  & 2.30 & 1.87 & 79	 & (1344, 1371) & 1914 & 0.30 & 814.0	& 4.4 & 31.2  & 0.2 \\
        aq & 0  & 1.72 & 1.36 & $-$85 & (1828, 1815) & 1915 & 0.22 &1018.6 & 5.1 & 71.7  & 0.4  \\
        ar & 40 & 2.38 & 1.78 & 84	 & (1389, 1423) & 1915 & 0.29 & 584.9	& 5.4 & 22.7  & 0.2 \\
        as & 0  & 1.98 & 1.33 & $-$76 & (2009, 2053) & 1915 & 0.20 & 582.0	& 5.3 & 36.4  & 0.3 \\
        b & 40  & 1.78 & 1.41 & 86	 & (1915, 1958) & 1915 & 0.21 & 2213.5	& 4.2 & 145.3 & 0.3  \\
        c & 0   & 2.73 & 1.79 & 87	 & (1435, 1461) & 1915 & 0.28 & 647.8	& 5.8 & 21.8  & 0.2 \\
        d & 30  & 1.94 & 1.55 & $-$63 & (2061, 1585) & 1917 & 0.23 & 1619.4& 5.5 & 88.7  & 0.3 \\
        e & $-$30 & 2.84 & 1.76 & $-$90 & (1442, 1472) & 1915 & 0.28&1147.9& 5.5 & 37.8  & 0.2 \\
        f & 0   & 2.89 & 1.73 & 90	 & (1782, 1468) & 1920 & 0.28 & 1016.9	& 5.6 & 33.5  & 0.2 \\
        g & 30  & 2.24 & 1.67 & 84	 & (1606, 1647) & 1915 & 0.25 & 1025.7	& 5.0 & 45.2  & 0.2 \\
        h & 0   & 1.73 & 1.27 & $-$84 & (1994, 2114) & 1912 & 0.21 & 620.9	& 4.8 & 46.6  & 0.4 \\
        i & 30  & 1.75 & 1.53 & $-$67 & (1682, 1641) & 1912 & 0.25 & 736.3	& 4.4 & 45.3  & 0.3 \\
        j & 30  & 2.83 & 1.68 & $-$88 & (1480, 1510) & 1915 & 0.27 & 970.3	& 5.9 & 33.6 & 0.2 \\
        k & 30  & 1.90 & 1.11 & $-$80 & (2227, 2207) & 1912 & 0.18 & 2283.9& 5.5 & 178.4& 0.4 \\
        l & 0   & 1.85 & 1.65 & $-$80 & (2419, 1525) & 1915 & 0.26 & 805.5	& 4.8 & 43.5 & 0.3 \\
        m & 30  & 2.57 & 2.12 & 82	 & (1373, 1377) & 1917 & 0.29 & 3644.8	& 4.5 & 110.2& 0.1 \\
        n & 30  & 1.85 & 1.14 & $-$82 & (1406, 1844) & 1914 & 0.19 & 580.8	& 5.6 & 45.4 & 0.4 \\
        o & 0   & 2.73 & 1.89 & $-$89 & (1310, 1252) & 1915 & 0.32 & 1503.	& 5.5 & 48.0 & 0.2 \\
        p & 0   & 1.70 & 1.23 & 86	 & (2078, 2057) & 1915 & 0.20 & 687.8	& 4.9 & 54.2 & 0.4 \\
        q & 0   & 2.43 & 1.86 & 88	 & (1345, 1367) & 1915 & 0.30 & 464.4	& 4.3 & 16.9 & 0.2 \\
        r & 30  & 3.16 & 1.90 & $-$83 & (1556, 1419) & 1917 & 0.29 & 820.8	& 5.1 & 22.5 & 0.1 \\
        s & 0   & 2.78 & 1.99 & 87	 & (1339, 1369) & 1917 & 0.30 & 1450.5	& 5.2 & 43.2 & 0.2 \\
        t & 0   & 2.31 & 1.95 & 74	 & (1297, 1293) & 1914 & 0.31 & 792.8	& 5.1 & 29.0 & 0.2 \\
        u & 30  & 2.06 & 1.21 & $-$75 & (1394, 2088) & 1914 & 0.19 & 729.5	& 6.0 & 48.2 & 0.4 \\
        v & 30  & 1.94 & 1.23 & $-$81 & (1410, 1916) & 1914 & 0.19 & 126.7	& 5.4 & 8.7  & 0.4 \\
        w & 30  & 2.47 & 1.74 & $-$82 & (1508, 1424) & 1915 & 0.28 & 680.6	& 4.6 & 26.1 & 0.2 \\
        x & 0   & 2.67 & 1.88 & $-$88 & (1321, 790)	& 1916 & 0.30 &514.1& 5.2 & 16.9 & 0.2 \\
        y & 0   & 2.34 & 1.78 & $-$86 & (1464, 1138) & 1917 & 0.27 & 378.3	& 5.5 & 15.0 & 0.2 \\
        z & 40  & 2.61 & 1.63 & 82	 & (1382, 1377) & 1915 & 0.29 & 617.2	& 5.3 & 23.9 & 0.2 \\
        \hline \hline
    \end{tabular}
\end{table*}

\begin{table*}
    \centering
    \caption{Global statistics for the 12-m SPW 27 cubes, after feathering data from the 12-m, 7-m, and TP arrays altogether, for all ACES fields. The noise is estimated from channels in the lowest quartile of the mean spectrum (assumed emission-free) using the scaled median absolute deviation (${\rm MAD}$; where $\sigma_{\rm MAD} = 1.4826\,{\rm MAD}$).}
    \label{tab:cubestats27}
    \begin{tabular}{cccccccccccc}
        \hline \hline
        Field & V$_{\textrm{off}}$ & B$_{\textrm{maj}}$ & B$_{\textrm{min}}$ & BPA & Image size & No.\ channels & Pixel size & S$_\text{peak}$ & $\sigma_{\rm MAD}$ & S$_\text{peak}$ & $\sigma_{\rm MAD}$ \\
        & (\kms) & (\arcsec) & (\arcsec) & (\arcdeg) & (pix) & & (\arcsec) & (\mjypbm{}) & (\mjypbm{}) & (K) & (K) \\
        \hline
        a & 40  & 1.74 & 1.19 & $-$77 & (2091, 2498) & 1912 & 0.20 & 89.1	& 4.7 & 7.0   & 0.4 \\
        aa & 0  & 2.08 & 1.73 & $-$75 & (1620, 1479) & 1917 & 0.26 & 160.4	& 4.6 & 7.2   & 0.2 \\
        ab & 0  & 1.97 & 1.69 & 89	 & (1494, 1259) & 1914 & 0.28 & 123.2	& 4.4 & 6.0   & 0.2 \\
        ac & 0  & 1.78 & 1.21 & $-$79 & (1683, 1624) & 1914 & 0.19 & 122.0	& 5.1 & 9.2   & 0.4 \\ 
        ad & 0  & 2.01 & 1.18 & $-$76 & (1182, 2562) & 1920 & 0.20 & 106.2	& 5.6 & 7.3   & 0.4 \\
        ae & 0  & 2.28 & 1.84 & 82	 & (1341, 1364) & 1914 & 0.30 & 126.0	& 4.7 & 4.9   & 0.2 \\
        af & 0  & 2.84 & 1.68 & $-$90 & (1443, 1174) & 1915 & 0.31 & 245.9	& 5.0 & 8.4   & 0.2 \\
        ag & 0  & 1.90 & 1.45 & $-$79 & (1729, 1719) & 1912 & 0.24 & 138.8	& 4.9 & 8.2   & 0.3 \\
        ah & 30 & 1.86 & 1.20 & $-$72 & (2001, 1939) & 1914 & 0.20 & 102.8	& 4.3 & 7.5   & 0.3 \\
        ai & 0  & 2.57 & 1.76 & 80	 & (1376, 1620) & 1914 & 0.29 & 161.2	& 4.6 & 5.8   & 0.2 \\
        aj &$-$30 & 2.42 & 1.77 & 84	 & (1438, 1286) & 1914 & 0.31 &172.1& 4.3 & 6.5   & 0.2 \\
        ak & 30 & 2.47 & 2.08 & 83	 & (762, 1007)	& 1916 & 0.32 & 173.6	& 4.7 & 5.5   & 0.1 \\
        al & 0  & 2.75 & 2.03 & 86	 & (2189, 1446) & 1920 & 0.28 & 187.6	& 5.6 & 5.4   & 0.2 \\
        am & 30 & 3.22 & 1.97 & $-$84 & (2610, 1138) & 1920 & 0.30 & 238.7	& 6.3 & 6.1   & 0.2 \\
        an & 30 & 2.52 & 1.72 & 83	 & (1333, 1357) & 1914 & 0.30 & 1114.7	& 5.0 & 41.7  & 0.2 \\
        ao & 0  & 2.52 & 1.62 & $-$87 & (2509, 921)	& 1914 & 0.28 &177.2& 5.9 & 7.0   & 0.2 \\
        ap & 0  & 2.27 & 1.86 & 79	 & (1342, 1369) & 1914 & 0.30 & 143.0	& 4.3 & 5.5   & 0.2 \\
        aq & 0  & 1.72 & 1.34 & $-$84 & (1826, 1813) & 1915 & 0.22 & 125.7	& 5.0 & 8.8   & 0.4 \\
        ar & 40 & 2.35 & 1.77 & 84	 & (1386, 1423) & 1914 & 0.29 & 157.5	& 5.3 & 6.1   & 0.2 \\
        as & 0  & 2.00 & 1.31 & $-$75 & (2005, 2051) & 1915 & 0.20 & 156.9	& 5.2 & 9.7   & 0.3 \\
        b & 40  & 1.76 & 1.40 & 85	 & (1913, 1956) & 1915 & 0.21 & 837.0	& 4.1 & 55.1  & 0.3  \\
        c & 0   & 2.74 & 1.78 & 87	 & (1433, 1459) & 1915 & 0.28 & 155.7	& 5.6 & 5.2   & 0.2 \\
        d & 30  & 1.96 & 1.53 & $-$65 & (2058, 1582) & 1917 & 0.23 & 105.6	& 5.4 & 5.7   & 0.3 \\
        e & $-$30 & 2.94 & 1.80 & 87	 & (1441, 1471) & 1915 & 0.28 &175.7& 5.5 & 5.4   & 0.2 \\ 
        f & 0   & 2.85 & 1.72 & 90	 & (1780, 1466) & 1920 & 0.28 & 222.3	& 5.4 & 7.4   & 0.2 \\
        g & 30  & 2.27 & 2.26 & 85	 & (1603, 1644) & 1915 & 0.25 & 193.6	& 4.8 & 6.1   & 0.2 \\
        h & 0   & 1.70 & 1.25 & $-$84 & (1991, 2112) & 1912 & 0.21 & 174.3	& 4.8 & 13.3  & 0.4 \\
        i & 30  & 1.73 & 1.52 & $-$70 & (1678, 1639) & 1912 & 0.25 & 141.8	& 4.4 & 8.7   & 0.3 \\
        j & 30  & 2.83 & 1.65 & $-$87 & (1477, 1508) & 1915 & 0.27 & 209.7	& 5.9 & 7.3  & 0.2 \\
        k & 30  & 1.89 & 1.09 & $-$79 & (2223, 2202) & 1912 & 0.18 & 113.5	& 5.5 & 8.9  & 0.4 \\
        l & 0   & 1.84 & 1.63 & $-$79 & (2417, 1522) & 1915 & 0.26 & 248.0	& 4.7 & 13.4 & 0.3    \\
        m & 30  & 2.55 & 2.10 & 81	 & (1372, 1376) & 1917 & 0.29 & 2547.4	& 4.4 & 77.1 & 0.1    \\
        n & 30  & 1.80 & 1.12 & $-$79 & (1403, 1842) & 1914 & 0.19 & 99.8	& 5.4 & 8.0  & 0.4 \\
        o & 0   & 2.82 & 1.91 & 86	 & (1308, 1251) & 1915 & 0.32 & 333.3	& 5.5 & 10.0 & 0.2 \\
        p & 0   & 1.82 & 1.35 & 84	 & (1887, 1868) & 1915 & 0.22 & 130.3	& 4.4 & 8.6  & 0.3 \\
        q & 0   & 2.41 & 1.85 & 88	 & (1343, 1365) & 1914 & 0.30 & 199.3	& 4.2 & 7.2  & 0.2 \\
        r & 30  & 3.25 & 1.91 & $-$86 & (1555, 1417) & 1917 & 0.29 & 204.3	& 5.0 & 5.3  & 0.1 \\
        s & 0   & 2.75 & 1.99 & 86	 & (1337, 1367) & 1917 & 0.30 & 229.0	& 5.0 & 6.8  & 0.1 \\
        t & 0   & 2.31 & 1.94 & 73	 & (1295, 1292) & 1914 & 0.31 & 149.8	& 4.9 & 5.4  & 0.2 \\
        u & 30  & 2.04 & 1.21 & $-$75 & (1392, 2085) & 1914 & 0.19 & 116.8	& 5.9 & 7.7  & 0.4 \\
        v & 30  & 1.92 & 1.22 & $-$81 & (1407, 1913) & 1914 & 0.19 & 115.1	& 5.3 & 8.0  & 0.4 \\
        w & 30  & 2.45 & 1.73 & $-$82 & (1506, 1423) & 1914 & 0.28 & 123.8	& 4.5 & 4.7  & 0.2 \\
        x & 0   & 2.64 & 1.86 & $-$88 & (1318, 790)	& 1916 & 0.30 &154.5& 5.2 & 5.1  & 0.2 \\
        y & 0   & 2.28 & 1.66 & $-$88 & (1462, 1136) & 1915 & 0.27 & 128.4	& 5.6 & 5.5  & 0.2 \\
        z & 40  & 2.59 & 1.63 & 82	 & (1380, 1375) & 1914 & 0.29 & 138.9	& 5.1 & 5.3  & 0.2 \\
        \hline \hline
    \end{tabular}
\end{table*}

\section{Array combination using Miriad}\label{app_sec:comb_miriad}
The 12-m array, the 7-m array, and the Total Power (TP) array of ALMA each are each only sensitive to a certain range of angular scales. To recover the multi-scale gas structures in the CMZ from 0.1~pc (2\arcsec{}) to over 200~pc, it is necessary to combine data from the different arrays and produce images that are sensitive to both small and large angular scales. Such a procedure, usually referred to as `array combination', can be implemented in different approaches, e.g., concatenating visibility data and CLEANing them to obtain an image (`joint imaging'), or linearly combining the images from different arrays (`feathering'). The ACES data releases took the latter approach and produced the combined images \citepalias{Ginsburg2026,Walker2026}. A detailed discussion on the array combination can be found in \citet{Plunkett2023}.

One of the reasons why we chose feathering over joint imaging is that joint imaging in CASA is computationally more expensive and slower, and often diverges. Meanwhile, array combination can be implemented in other packages, e.g., Miriad \citep{Stanimirovic1999,Stanimirovic2002}.

Here, we present an experiment of joint imaging using Miriad, on the SiO (2--1) line in one of the regions that cover the G0.253$-$0.016 cloud. The joint imaging is implemented with a modified ALMICA procedure\footnote{\url{https://github.com/xinglunju/almica}}, which is forked from the original version\footnote{\url{https://github.com/baobabyoo/almica}} by Hauyu Baobab Liu \citep{Liu2013}. The workflow consists of the following major steps:

\begin{enumerate}
    \item Deconvolve the TP image to obtain a CLEAN model for each of the 7-m pointings, using the \texttt{CLEAN} and \texttt{DEMOS} tasks.
    \item Generate artificial visibility data from the TP CLEAN model, using the \texttt{UVRANDOM} and \texttt{UVMODEL} tasks. Adjust the weight of the TP visibility data through manipulating the system temperature, using the \texttt{UVPUTHD} task.
    \item Image the TP and 7-m visibilities together to obtain a CLEAN model for the whole mosaic, using the \texttt{INVERT} and \texttt{MOSSDI} tasks.
    \item Decompose the TP+7-m CLEAN model for the mosaic to that for each of the 12-m pointings, using the \texttt{DEMOS} task. Generate artificial visibility data from the CLEAN model for individual 12-m pointings, using the \texttt{UVRANDOM} and \texttt{UVMODEL} tasks. Adjust the weight of the TP+7-m visibility data through manipulating the system temperature, using the \texttt{UVPUTHD} task.
    \item Finally, image the TP+7-m and 12-m visibilities together to obtain the combined image, using the \texttt{INVERT}, \texttt{MOSSDI}, and \texttt{RESTOR} tasks.
    \item There is an optional step to further merge the TP image with the combined image, using the \texttt{IMMERGE} task. This linear combination in the image domain is similar to the feathering method in CASA.
\end{enumerate}
For detailed steps, readers are referred to Appendix~A of \citet{Liu2013} and the ALMICA github repository.

\begin{figure*}
\centering
\includegraphics[width=1.0\textwidth]{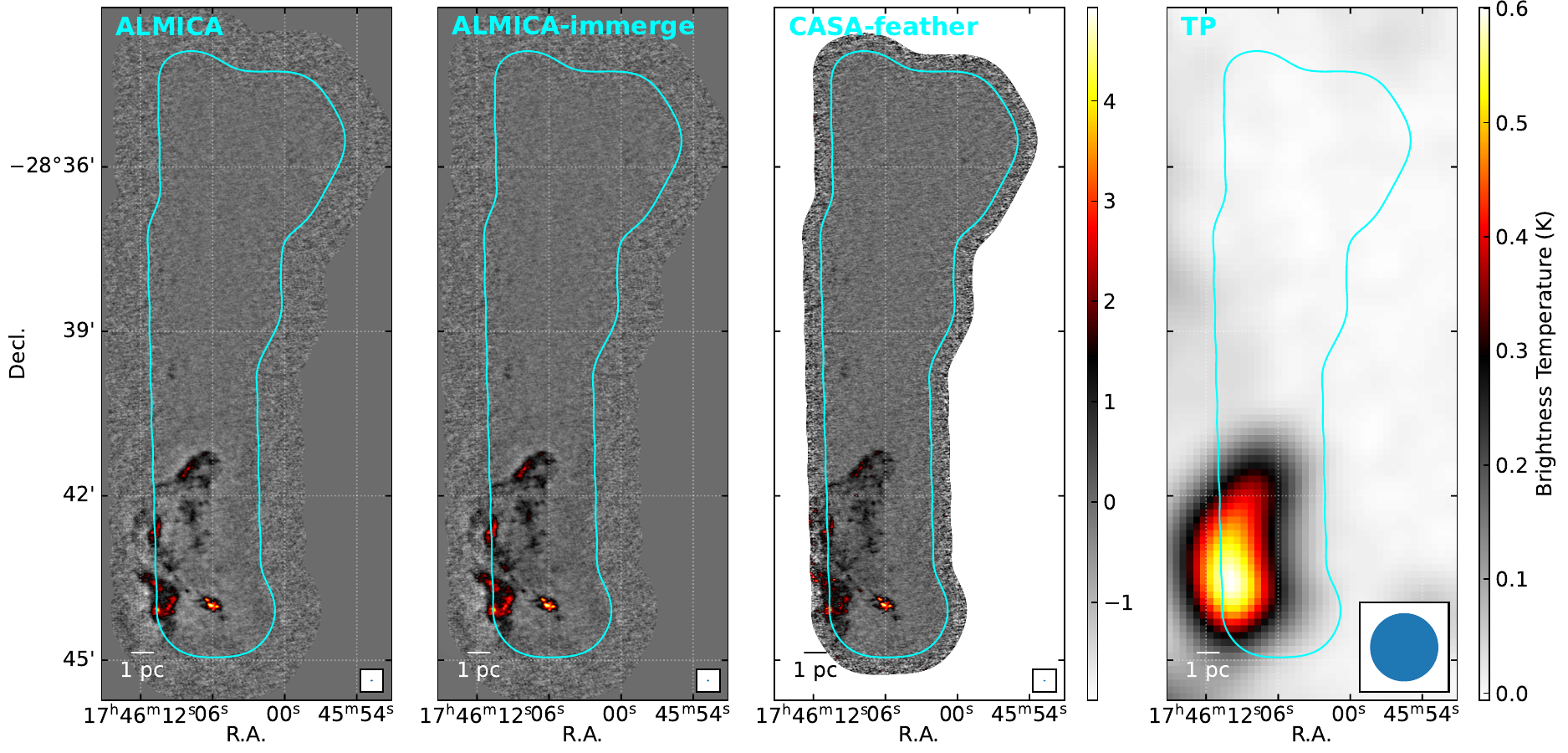}
\caption{SiO images of a velocity channel at $\sim$30~\kms{} from different array combination approaches. The velocity is chose to be around the systemic velocity of the cloud and show complex structures in emission. The three array-combined images are adjusted to the same color scale. The TP image with a different color scale is also shown for reference. Primary-beam response at the 0.5 level of the 12-m array mosaic is shown as a cyan contour in all panels.
}
\label{fig:arraycomb_onechan}
\end{figure*}

\begin{figure*}
\centering
\includegraphics[width=1.0\textwidth]{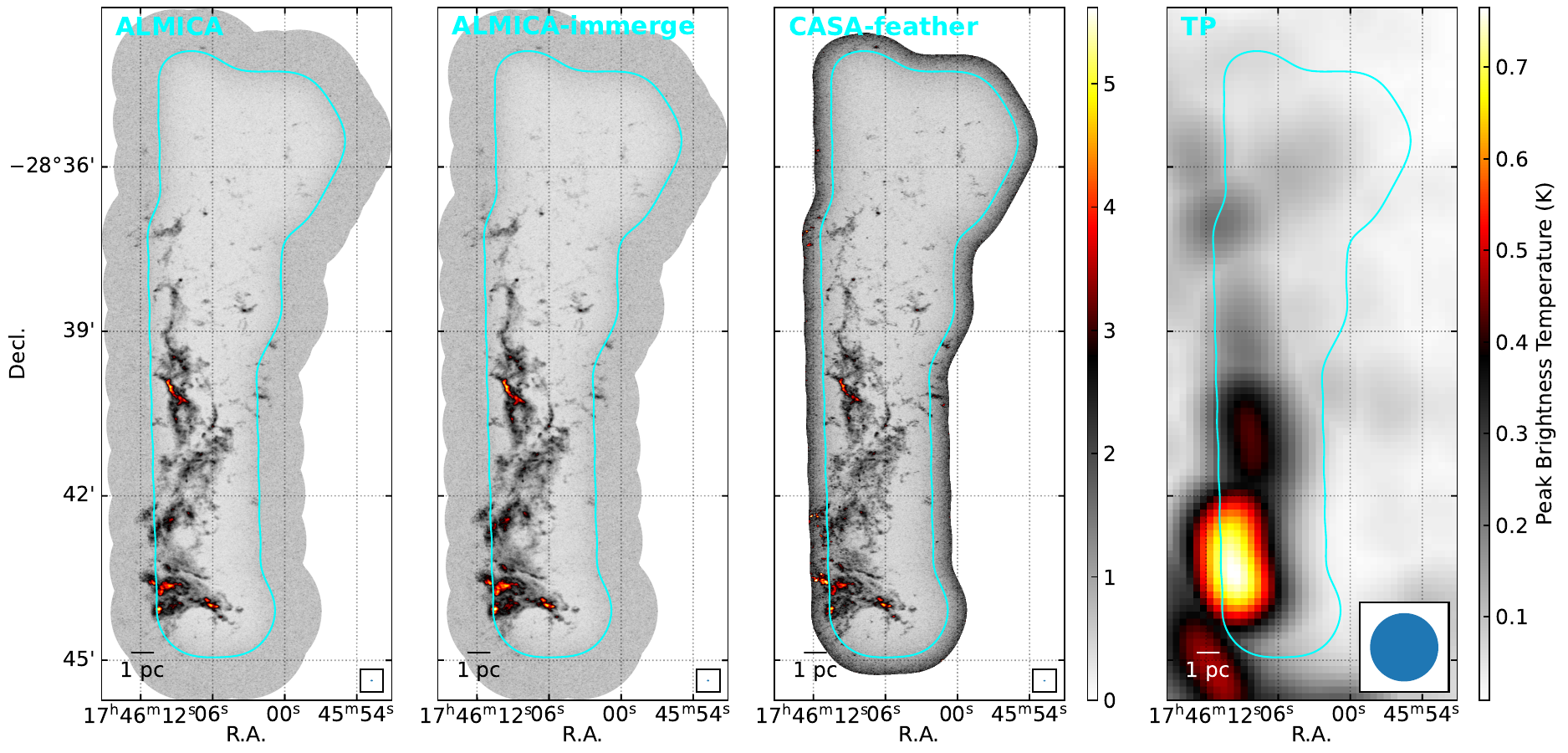}
\caption{Peak brightness temperatures of the SiO (2--1) line, from different array combination approaches with the same color scale, and from the TP array with a different color scale for reference.}
\label{fig:arraycomb_mom8}
\end{figure*}

\begin{figure*}
\centering
\includegraphics[width=1.0\textwidth]{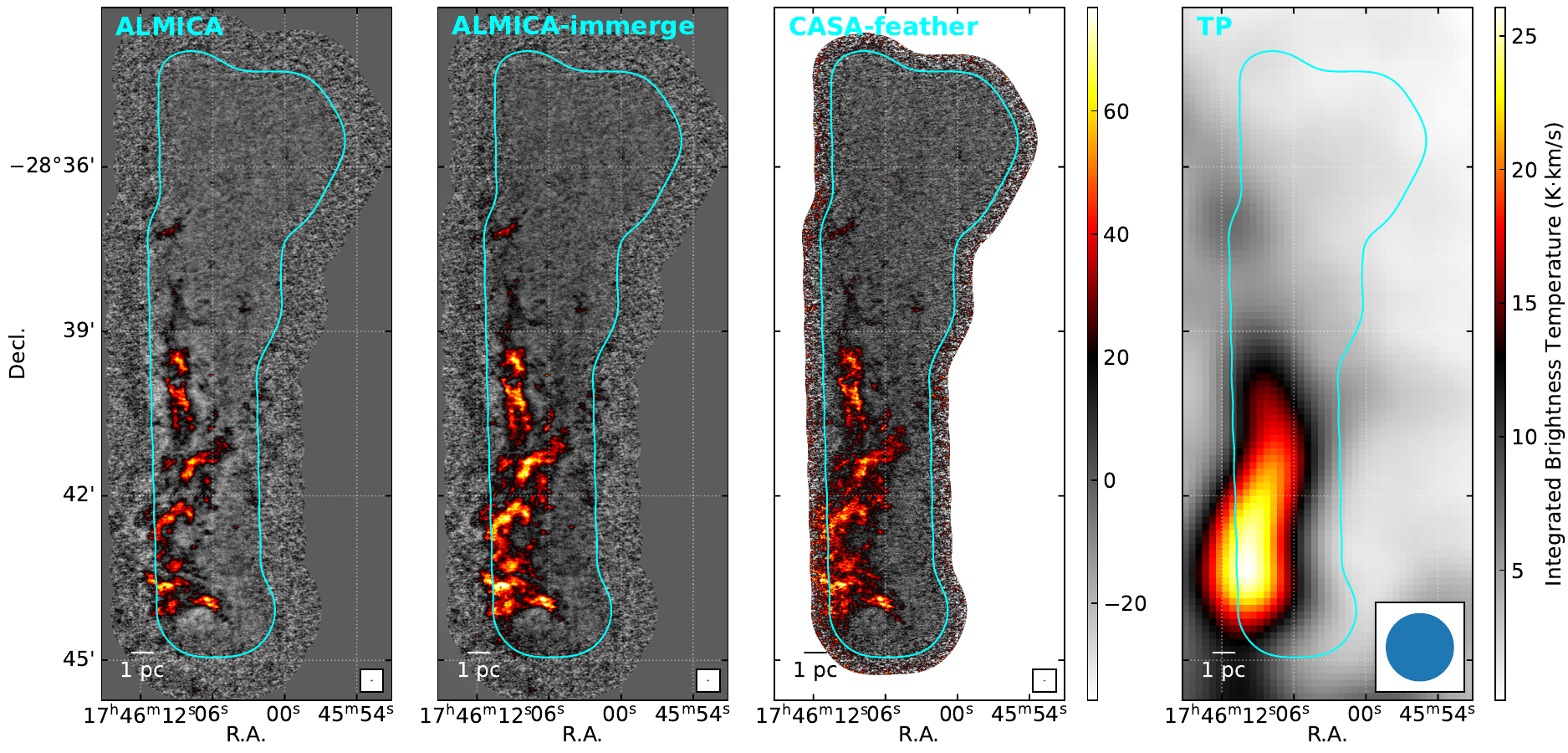}
\caption{Integrated brightness temperatures of the SiO (2--1) line, from different array combination approaches with the same color scale, and from the TP array with a different color scale for reference. The images are integrated between $\vlsr{}=-100$ and $200$~\kms{}.}
\label{fig:arraycomb_mom0}
\end{figure*}

We follow the above workflow, and set a system temperature of 500~K for both the TP and the 7-m visibilities. In step 2, we generate 500 data points for the artificial TP visibility data, which are uniformly distributed in the $uv$ plain within a maximum baseline length of 12~m. Similarly, in step 4, we generate 3000 data points for the artificial TP+7-m visibility data uniformly distributed in the $uv$ plain within a maximum baseline length of 45~m. The resulting combined images with and without the optional \texttt{IMMERGE} step are shown in Figures~\ref{fig:arraycomb_onechan}--\ref{fig:arraycomb_mom0}. We also show the combined image from the feathering approach and the TP image for comparison. The beam sizes and image RMS of the three approaches are summarised in \autoref{tab:combinedimages} .

It can be seen that the combined images made with ALMICA and CASA-feather are broadly consistent in terms of the morphology, although some nuances can still be found. The ALMICA image clearly shows negative bowls around bright emission peaks, which are signatures of missing flux \citep{Plunkett2023}. With the additional \texttt{IMMERGE} step, ALMICA imaging performs better in terms of mitigating the negative bowls. The CASA-feather image performs the best in recovering diffuse emission, where the negative bowls are mostly gone.

\begin{table*}
\centering
\caption{Imaging parameters and properties of the combined SiO (2--1) images from ALMICA and CASA-feathering.\label{tab:combinedimages}}
\begin{tabular}{ccccc}
\hline
Combination method & Beam size \& PA & Image RMS &  \\
 & (\arcsec{} $\times$ \arcsec{}, \arcdeg{}) & (\mjypbm{} per 0.85~\kms{}) \\
\hline
Miriad-ALMICA         & 2.58 $\times$ 1.91, $-$88.24 & 4.0 \\
Miriad-ALMICA-immerge & 2.58 $\times$ 1.91, $-$88.24 & 4.0 \\
CASA-feathering       & 2.52 $\times$ 1.62, $-$86.95 & 4.2 \\
\hline 
\end{tabular}
\end{table*}

\begin{figure*}
\centering
\includegraphics[width=1.0\textwidth]{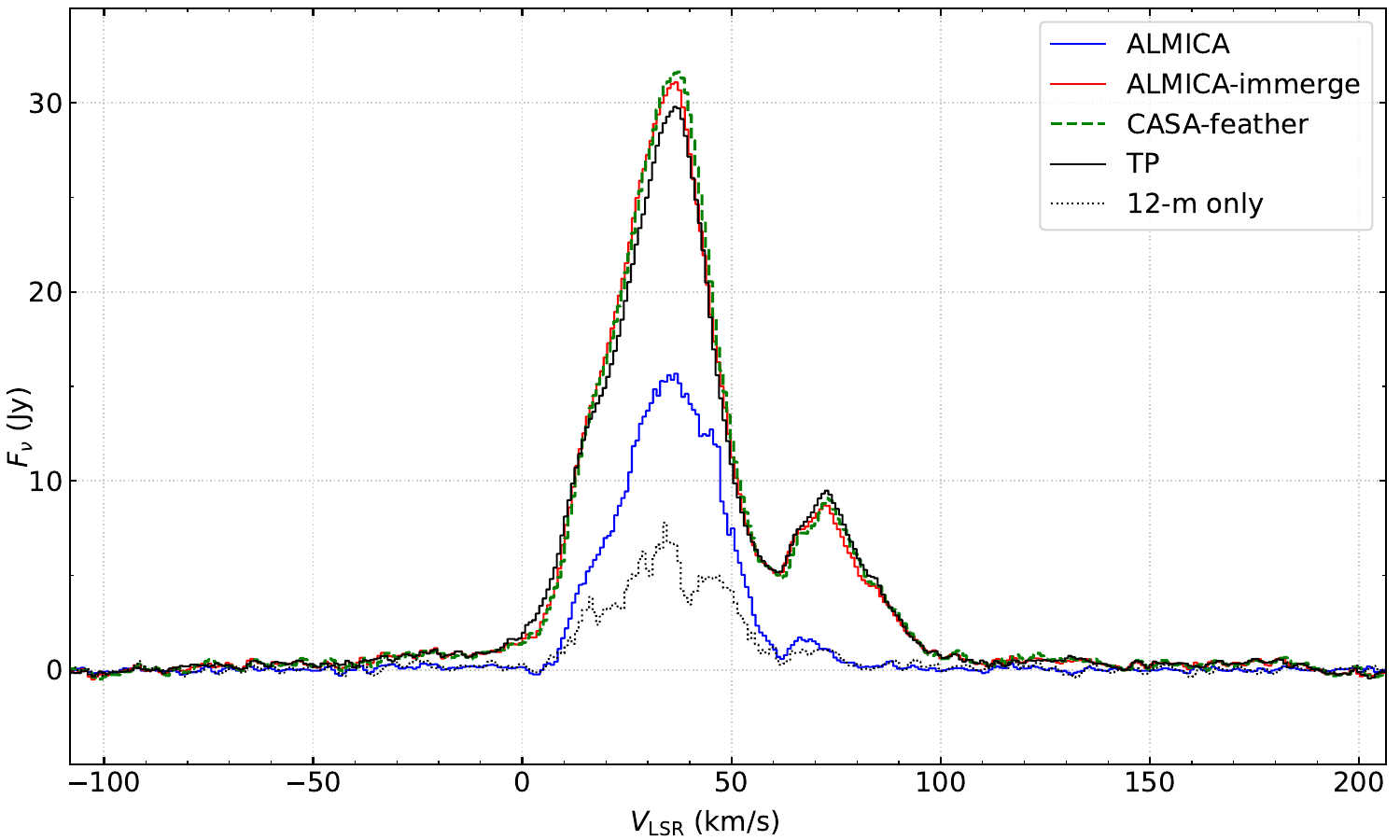}
\caption{SiO (2--1) spectra with flux densities in the y-axis within a circle of a radius of 74\farcs{8}, which is the beam size of the TP image, centered at the peak of the integrated brightness temperature map (RA, Dec.\ = 17:46:06.57, $-$28:43:52.12 in the ICRS frame). Spectra from the three different array combination approaches, as well as those from the TP image and the 12-m array only image are also plotted for references.}
\label{fig:arraycomb_spec}
\end{figure*}

In \autoref{fig:arraycomb_spec} we compare spectra from different array combination approaches. Spectra from the TP image and from the 12-m array only image (made with the CASA \textit{tclean} task) are also plotted for comparison. The ALMICA image clearly recovers more flux than the 12-m array only one. However, it does not recover all of the expected flux based on the TP image. The ALMICA image with the additional \texttt{IMMERGE} step and the CASA-feather image perform better when it comes to recovering the flux of the TP image.

To summarize, the experiment suggests that the ALMICA approach with the additional \texttt{IMMERGE} step performs equally well at mitigating negative bowls and recovering the missing flux as compared to the CASA-feather approach, but at a computing speed of about 10 times faster. The drawbacks of ALMICA, however, include its reliance on the Miriad package, which has a smaller user community than CASA, and its current lack of parallelisation.

\clearpage

\bsp
\label{lastpage}
\end{document}